# SURFACE OBSERVABLES, 2-KNOT INVARIANTS, AND NONABELIAN ELECTRIC FLUXES

ALBERTO S. CATTANEO

ABSTRACT. This work introduces a surface observable for nonabelian four-dimensional $BF$ theory with a cosmological term. The surface observable yields new 2-knot invariants that may extend beyond known examples such as the Alexander invariant. By BV pushforward, the surface observable induces an electric observable in nonabelian Yang–Mills theory, offering a concrete realization of 't Hooft operators. An application to self-dual Yang–Mills theory is also discussed.

## Contents

2020 *Mathematics Subject Classification.* 81T70, 81T13, 57R56.

I acknowledge partial support of the SNF Grant No. 200021_227719 and of the Simons Collaboration on Global Categorical Symmetries. This research was (partly) supported by the NCCR SwissMAP, funded by the Swiss National Science Foundation. This article is based upon work from COST Action 21109 CaLISTA, supported by COST (European Cooperation in Science and Technology) (www.cost.eu), MSCA-2021-SE-01-101086123 CaLIGOLA, and MSCA-DN CaLiForNIA -101119552.

1. Introduction

The main goal of this paper is to introduce a surface observable for $BF$ theory with cosmological term in four dimensions and to use it to define 2-knot invariants and an electric observable for nonabelian Yang–Mills theory. We shortly recapitulate the two independent parts.

1.1. **The electric observable.** Recall that in abelian Yang–Mills theory (i.e., in electromagnetism), the curvature $F_A = \mathrm{d}A$ is gauge invariant. As a result, we can consider observables like $\int_\Sigma F_A$ (magnetic flux) and $\int_\Sigma *F_A$ (electric flux). Here $*$ is the Hodge dual by the background metric (which in this paper we assume to be riemannian).

If $\Sigma$ has a boundary $\gamma$, then, in the abelian case, $\int_\Sigma F_A = \int_\gamma A$, so $\exp\left(\mathrm{i}\lambda \int_\Sigma F_A\right)$ may be understood as a Wilson loop along $\gamma$, which makes sense also in the nonabelian case. If $\Sigma$ is closed, a nonabelian version of $\exp\left(\mathrm{i}\lambda \int_\Sigma F_A\right)$ was introduced in [2].

The goal of this paper is to define a nonabelian electric observable, by correcting the abelian one by suitable $\hbar$-corrections:
$$\mathcal{O}_\Sigma = \mathrm{e}^{\frac{\mathrm{i}}{4\lambda} \int_\Sigma \langle \xi, *F_A \rangle + \cdots}.$$

Here $\lambda$ is related to the Yang–Mills coupling constant, $\langle\ ,\ \rangle$ is an invariant pairing on the Lie algebra (e.g., the Killing form if we work with a semisimple, compact Lie group), and $\xi$ is a chosen element of the Lie algebra. In this paper, we work perturbatively, so $\xi$ can be anything. However, we expect that $\xi$ should eventually belong to a prequantizable coadjoint orbit.

These electric observables may be an explicit realization of those proposed by 't Hooft [67] which play a preeminent role in the modern relation between topological operators and symmetry [44, 41] and may be regarded as topological defects (see, e.g., among others [39, 43, 5, 15, 59] and references therein).

1.2. **The surface observable and 2-knot invariants.** The starting point of our construction is a surface observable for $BF$ theory with cosmological term. This is a generalization of the surface observable for pure $BF$ theory introduced in [31]. As the latter, this observable is defined in terms of additional surface fields. Related constructions appear in [8, 58] and in [56, 2, 60]. Unlike the observables proposed in [35, 23, 21], which required to view the surface as a loop of loops, this observable is intrinsically defined in terms of the surface with no additional structure.

Its explicit expression as well as its expectation value can be written in terms of integrals over compactified configuration spaces à la Fulton–MacPherson–Axelrod–Singer [40, 6, 7] and yields an invariant up to isotopy of the embedded surface (in this paper, like in [31], we focus on the case of a long 2-knot in $\mathbb{R}^4$, but the construction is more general).

In the pure case, the invariants produced in [31, 62] contain, as a special case, the one proposed by Bott in [16], have been proved by Watanabe [68] (see also [49, 50]) to be related to the Alexander invariant, and should be related to the invariants introduced by Bar-Natan in [10]. The new invariants introduced in this paper might be independent and deserve further study.

1.3. **The BV formalism and the BV pushforward.** The main technique to produce the surface observable, as a quantum Batalin–Vilkovisky (BV) observable, is a technique, pioneered in [31] and [51] (see also [54, 55, 33, 26, 34, 27]), called BV pushforward. It allows for integrating out some additional fields, like the surface fields we introduce to define the surface observable.

The same technique may be used to pass from one field theory to another (often producing an equivalence). In this paper, we rely on the equivalence (in the appropriate sense) between four-dimensional $BF$ theory with cosmological term and four-dimensional Yang–Mills theory (with quantum corrections) devised in [13]. (This is a sort of four-dimensional version of the exact relationship between $BF$ and Yang–Mills theory in two dimensions [70].)

This construction allows us to turn the surface observable for $BF$ theory into an electric observable for Yang–Mills theory. A similar construction allows for the construction of surface observables in self-dual Yang–Mills theory.

The surface observable constructed in this paper might ultimately also allow us to construct a surface observable for four-dimensional gravity, using results of [25] and references therein. We defer this to future work.

1.4. **Plan of the paper.** In Section 2 we recall classical $BF$ theory with cosmological term in four dimensions and heuristically construct a surface observable. In Section 3, we recall the basics of the BV formalism and of the BV pushforward. This section may be skipped by the reader already familiar with these topics. Section 4 is the core of the paper: there we introduce the surface observable and discuss its properties. In Section 5, we describe the Feynman-diagram integration of the surface fields, producing a surface observable for $BF$ theory as a functional of the $BF$-theory BV fields only. Section 6 contains the first application: the construction of 2-knot invariants. Section 7 contains the second application: the construction of an electric observable for nonabelian Yang–Mills theory. Finally, Section 8 applies the construction to the simpler case of self-dual Yang–Mills theory.

**Acknowledgements.** I thank Pavel Mnev, Gregory Moore, and Nima Moshayedi for useful discussions.

## 2. The classical theory

In this section we analyze the classical theory and its surface observable, showing that it is well-defined on shell. This is intended to give an intuitive understanding of the matter which will be refined and made precise via the BV formalism in the next sections.

2.1. **The action.** Our theory is $BF$ theory in four dimensions. The data are an oriented four-manifold[1] $M$ and a $G$-principal bundle $P \to M$. The fields are a connection 1-form $A$ and a 2-form $B$ in the adjoint bundle ad$P$. In the simplified case when $P$ is trivial, we view $A$ and $B$ as elements of $\Omega^1(M) \otimes \mathfrak{g}$ and of $\Omega^2(M) \otimes \mathfrak{g}$, respectively, where $\mathfrak{g}$ is the Lie algebra of $G$. Finally, we assume that the Lie algebra $\mathfrak{g}$ is unimodular and is endowed with a nondegenerate invariant pairing $\langle\,,\,\rangle$.

Pure $BF$ theory is described by the action functional $S[A, B] := \int_M BF_A$, where $F_A$ is the curvature of $A$ ($F_A = \mathrm{d}A + \frac{1}{2}[A, A]$), and the pairing between the sections of the adjoint bundle (or of the $\mathfrak{g}$-valued fields) is understood (we prefer to avoid the more pedantical notation $\langle B, F_A \rangle$ instead of $BF_A$).[2]

---

[1] If the manifold is not compact, suitable vanishing conditions of the fields are assumed.

In this paper we use an extension of pure $BF$ theory involving the "cosmological term" (so called because of its relation with the cosmological term in gravity) $\langle B, B \rangle$, which we will simply write $BB$:

$$(2.1) \qquad S[A, B] := \int_M \left( BF_A - \frac{\lambda}{2} BB \right),$$

where $\lambda$ is a parameter (with $\lambda = 0$ we recover the pure theory).

*Remark* 2.1. From now on, whenever two ad-sections (or $\mathfrak{g}$-valued fields) are juxtaposed, the pairing $\langle \ , \ \rangle$ is understood.[3]

*Remark* 2.2. The minus sign in front of $\lambda$ is purely conventional (we prefer to put a minus sign here so as to have a plus sign in (2.4) instead of the other way around).

By taking a variation of the action, we see that the Euler–Lagrange (EL) equations are

$$(2.2) \qquad F_A = \lambda B \quad \text{and} \quad d_A B = 0.$$

Note that, by the Bianchi identity, the second equation is redundant as long as $\lambda \neq 0$.

The theory has the infinitesimal symmetries

$$(2.3a) \qquad \delta A = -d_A \gamma - \lambda \theta,$$
$$(2.3b) \qquad \delta B = -d_A \theta + [\gamma, B],$$

where the generators $\gamma$ and $\theta$ are ad-valued 0- and 1-forms, respectively. (Also here the choice of signs is purely conventional. It is made in such a way that the signs will disappear in (4.3) below.)

A classical observable is a function of the fields $A$ and $B$ that is invariant under the transformations (2.3). An on-shell observable is a function that is invariant upon using the EL equations (2.2).

---

[2]The invariant pairing is actually not needed in this setting, for we can define $B$ to be a 2-form in the coadjoint bundle $\mathrm{ad}^* P$ (or, in the simplified case when $P$ is trivial, as an element of $\Omega^2(M) \otimes \mathfrak{g}^*$). In this case, $BF_A$ is a shorthand notation for $(B, F_A)$, where $( \ , \ )$ is the canonical pairing. We prefer to use the version with the invariant pairing because it fits the extension we are going to consider next.

[3]If $\mathfrak{g}$ is semisimple compact, we may take $\langle \ , \ \rangle$ to be the Killing form. In this case, the convention is just that the trace symbol is understood (or defined to be part of the integral sign $\int_M$).

2.2. **The surface observable.** Consider an oriented 2-manifold $\Sigma$ and a map $\sigma\colon \Sigma \to M$. (In the quantum theory, $\sigma$ will have to be an embedding to avoid singularities in the expectation value of the observable we are going to introduce; however, at this level we can work with any map $\sigma$.)

We can use $\sigma$ to pull back the adjoint bundle ad$P$ as well as the fields $A$ and $B$. For simplicity of notation, we will keep writing $A$ and $B$ instead of $\sigma^*A$ and $\sigma^*B$. We now introduce new fields on $\Sigma$, both taking values in $\sigma^*\text{ad}P$: a 1-form $\alpha$ and a 0-form $\beta$. We then consider the following version of two-dimensional $BF$ theory coupled to $A$ and $B$ as external sources:

$$(2.4) \qquad J[\alpha,\beta;A,B] := \int_\Sigma \left(\beta\,\mathrm{d}_A\alpha + \frac{\lambda}{2}\beta[\alpha,\alpha] + \beta B\right).$$

Next we formally define the functional

$$(2.5) \qquad \mathcal{O}_\sigma[A,B] := \int D\alpha\, D\beta\, \mathrm{e}^{\frac{\mathrm{i}}{\hbar}J[\alpha,\beta;A,B]}.$$

We claim this is an observable. In fact,

$$\delta J = \int_\Sigma \left(-\beta[\mathrm{d}_A\gamma,\alpha] + \beta[\gamma,B] - \lambda\beta[\theta,\alpha] - \beta\mathrm{d}_A\theta\right),$$

but this can be compensated by the infinitesimal transformation

$$\alpha \rightsquigarrow \alpha - \theta - [\gamma,\alpha], \quad \beta \rightsquigarrow \beta - [\gamma,\beta],$$

in the integration variables $\alpha$ and $\beta$. Under the assumption that $\mathfrak{g}$ is unimodular, this transformation formally leaves the functional measure invariant.

As $\mathcal{O}_\sigma$ is an observable, we can consider its expectation value in the ambient four-dimensional $BF$ theory. As the latter is topological, this will produce an invariant of $\sigma\colon \Sigma \to M$. (It is at this point that we should insist that $\sigma$ is an embedding.)

We now turn to the problem of making some sense of the functional integral (2.5). Even if we define it perturbatively, the integral is ill-defined because of the "on-shell symmetry"

$$\widehat{\delta}\alpha = -\mathrm{d}_A p - \lambda[\alpha,p], \quad \widehat{\delta}\beta = \lambda[p,\beta],$$

where $p$ is a 0-form in $\sigma^*\text{ad}P$. In fact we have

$$\widehat{\delta}J = \int_\Sigma [\beta,p](F_A - \lambda B).$$

This means that, upon using the EL equations (2.2) (i.e., precisely around where the functional integral in $A$ and $B$ will be localized), the transformation $\widehat{\delta}$ is a symmetry for $J$. To make the functional integral in $\alpha$ and $\beta$ well-defined so as to make sense of (2.5), we have to impose a gauge fixing. The problem is however subtle because, as soon as we are off shell in the ambient theory, the transformation $\widehat{\delta}$ ceases to be a symmetry, yet we are still gauge fixing it. The right way to deal with this problem is to resort to the BV formalism and to replace (2.5) by a BV pushforward.

## 3. The BV formalism and the BV pushforward

In this section, we recall the basic facts about the BV formalism [12] that we are going to use, especially the BV pushforward technique. This cannot be an exhaustive introduction, for which we refer to [29, 28] and references therein (in particular, [27, 57]), but is just a short review also aimed to fix our notations. Readers familiar with the formalism can easily skip this section.

3.1. **The classical BV formalism.** The BV formalism is based on a triple $(\mathcal{F}, \omega, \mathcal{S})$, where $\mathcal{F}$ is a supermanifold, $\omega$ an odd symplectic form, and $S$ an even function satisfying the Classical Master Equation (CME)
$$(\mathcal{S}, \mathcal{S}) = 0,$$
where ( , ) (the BV bracket) is the odd Poisson bracket induced by $\omega$. It is useful to introduce the hamiltonian vector field $Q$ of $S$, called the BV operator,

(3.1) $$\iota_Q \omega = \delta \mathcal{S},$$

where $\iota$ denotes contraction, and $\delta$ is the exterior derivative on $\mathcal{F}$. Equivalently, we have $Qf = -(S, f)$ for every function $f$ (the minus sign is also conventional).[4] Note that $(\mathcal{S}, \mathcal{S}) = -Q\mathcal{S}$, so the CME is equivalent to $Q\mathcal{S} = 0$. Moreover, it implies that $[Q, Q] = 0$.

It is convenient to assume that $\mathcal{F}$ has an additional $\mathbb{Z}$-grading (roughly speaking, local coordinates are assigned an integer), which we call the ghost number gh in accordance with the applications in physics. In this setting, we require $\mathrm{gh}\,\mathcal{S} = 0$ and $\mathrm{gh}\,\omega = -1$. This implies $\mathrm{gh}\,Q = 1$, so $Q$ is a differential on $C^\infty(\mathcal{F})$ (it is therefore called a cohomological vector field).

In the application to field theory, $\mathcal{F}$ is infinite-dimensional, so some care has to be taken. Typically we assume $\omega$ to be a weak symplectic form, meaning that a function has a unique hamiltonian vector field if any. Therefore, we have to assume that $\mathcal{S}$ has a hamiltonian vector field.

---

[4]Our convention is to define the hamiltonian vector field $X_g$ of a function $g$ by $\iota_{X_g}\omega = \delta g$ and the BV bracket by $(g, h) = (-1)^{g+1} X_g h$.

The passage from a classical field theory to its BV version goes through extending the space of fields $F$ by adding to it the generators of the infinitesimal (on-shell) symmetries regarded as odd variables of ghost number one (the ghosts). If the symmetries are reducible (as is the case, e.g., in four-dimensional $BF$ theory), ghosts-for-ghosts (of increasing ghost number and alternating parity) have also to be included. We denote this extended space of fields by $\mathcal{M}$. The BV manifold will then be $\mathcal{F} = T^*[-1]\mathcal{M}$ with its canonical symplectic form (this notation means that the momenta—the coordinates of the cotangent fiber—have opposite parity to the corresponding base coordinate and ghost number equal to the opposite of that of the base coordinate minus one). The BV action $\mathcal{S}$ has the form $\mathcal{S} = S + \cdots$, where $S$ is the classical action and the dots denote terms depending on the momenta (i.e., vanishing when restricted to $\mathcal{M}$). The linear term in the momenta of the classical fields is responsible for the infinitesimal (on-shell) symmetries. Namely, if $\phi$ is a classical field, $Q\phi$ is, up to terms in the momenta, the infinitesimal symmetry parametrized by the ghosts.

One notable feature of the BV formalism is that $Q$ applied to the momentum of a classical field starts with the variation of $S$ with respect to that field. That is, the BV formalism knows not only of the symmetries but also of the EL equations. In particular, it may combine the two and deal, in a consistent way, with symmetries that exist only on shell.

3.2. **The quantum BV formalism.** The BV formalism is introduced to formalize the independence of the choice of gauge fixing. This is motivated by the following facts (see [20] and references therein, in particular:[12, 64, 47, 45, 46, 65]) which hold for a finite-dimensional, oriented, odd symplectic manifold $(\mathcal{F}, \omega)$:

   **(BV1):** A half-density[5] on $\mathcal{F}$ becomes a density when restricted to a lagrangian submanifold, so it can be integrated there (if the integral converges).
   **(BV2):** There is a canonical odd differential operator $\triangle$ (in the $\mathbb{Z}$-version, $\triangle$ has ghost number 1) called the BV Laplacian that acts on half-densities and satisfies $\triangle^2 = 0$ and the following properties:
     (1) $\int_{\mathcal{L}} \triangle \rho = 0$ for every lagrangian submanifold $\mathcal{L}$ and every half-density $\rho$ which is integrable over $\mathcal{L}$.

---

[5]A half-density is locally a function but transforms from chart to chart with the square root of the Berezinian of the transformation. As a consequence, its square can be integrated, up to convergence issues.

(2) If $\triangle \rho = 0$ and $\mathcal{L}_t$ is a family of lagrangian submanifolds, with $\rho$ integrable on every $\mathcal{L}_t$, depending smoothly on a parameter $t$, then $\frac{\mathrm{d}}{\mathrm{d}t} \int_{\mathcal{L}_t} \rho = 0$.

*Remark* 3.1. The above can be partially understood in terms of ordinary differential geometry for $\mathcal{F} = T^*[-1]M$ with $M$ an ordinary oriented manifold [69, 64]. Half-densities on $T^*[-1]M$ can be canonically identified with differential forms on $M$. Under this identification, $\triangle$ corresponds to the exterior derivative. Stokes theorem then yields the properties of $\triangle$ in the special case when the lagrangian submanifolds are of the form $N^*[-1]C$, where $C$ is a submanifold of $M$ and $N^*$ denotes the conormal bundle. The result for general lagrangian submanifolds is the discovery of Batalin and Vilkovisky [12].

Usually one prefers to work with functions instead of half-densities. To do so, one chooses[6] a reference, nowhere vanishing half-density $\rho$ satisfying $\triangle \rho = 0$ and defines the BV Laplacian on functions by $\Delta f \coloneqq \triangle(f\rho)/\rho$. The BV Laplacian on functions clearly satisfies $\Delta^2 = 0$. More surprisingly, independently of $\rho$, we have

$$\Delta(fg) = (\Delta f)g + (-1)^f f \Delta g - (-1)^f (f, g),$$

where $(-1)^f = \pm 1$ denotes the parity of $f$ and $(\ ,\ )$ is the BV bracket. In the case of the exponential of an even function $f$, one has

(3.2) $$\Delta \mathrm{e}^f = \left(\Delta f - \frac{1}{2}(f, f)\right) \mathrm{e}^f.$$

3.3. **BV functional integrals.** The application to physics is the following setting. One starts with a classical action $S$ on a space $F$ of classical fields. The goal is to make sense of the functional integral $\int_F \mathrm{e}^{\frac{\mathrm{i}}{\hbar}S}$. If the action has symmetries, the critical points are degenerate, so the saddle point approximation cannot be used. One then introduces the space $\mathcal{G}$ of generators of the infinitesimal symmetries and considers instead the integral $\int_{F \times \mathcal{G}[1]} \mathrm{e}^{\frac{\mathrm{i}}{\hbar}S}$. Since we have made the generators odd via the shift in $\mathcal{G}[1]$ (they are now the ghosts), we have to compute a berezinian integral, which vanishes because the classical action does not depend on the ghosts. Therefore, the integral is of the form infinity times zero—undetermined but already better than just infinity.

---

[6]The existence of such a $\rho$ in the finite-dimensional case is a consequence of the fact that an odd symplectic manifold is always symplectomorphic, although in a noncanonical way, to an odd cotangent bundle, $\mathcal{F} \cong T^*[-1]M$ with $M$ an ordinary oriented manifold [64]. Any volume form on $M$ provides such a reference half-density.

The next step is to consider the odd symplectic manifold $\mathcal{F} := T^*[-1](F \times \mathcal{G}[1])$, in which $\mathcal{L}_0 := F \times \mathcal{G}[1]$ sits as a lagrangian submanifold, and to extend the classical action to a functional $\mathcal{S}$ such that

(1) $\mathcal{S}|_{\mathcal{L}_0} = S$, and
(2) $\Delta e^{\frac{i}{\hbar}\mathcal{S}} = 0$.

We now consider a family $\mathcal{L}_t$ of lagrangian submanifolds deforming $\mathcal{L}_0$ such that the critical points of $\mathcal{S}$ on $\mathcal{L}_t$ are nondegenerate for $t \neq 0$. We then replace the ill-defined integral $\int_{\mathcal{L}_0} e^{\frac{i}{\hbar}\mathcal{S}}$ with $\int_{\mathcal{L}_t} e^{\frac{i}{\hbar}\mathcal{S}}$, $t \neq 0$, where the saddle-point approximation now makes sense. By the BV property (BV2.2), it does not matter which $\mathcal{L}_t$, $t \neq 0$, we choose. Such an $\mathcal{L}_t$ is called a gauge-fixing Lagrangian.[7]

The main problem in the infinite-dimensional case is that the BV Laplacian does not make sense as it is: it can be properly defined only after a regularization is introduced.

However, by (3.2), the condition $\Delta e^{\frac{i}{\hbar}\mathcal{S}} = 0$ is equivalent to the Quantum Master Equation (QME)

$$\frac{1}{2}(\mathcal{S}, \mathcal{S}) - i\hbar \Delta \mathcal{S} = 0.$$

Here we see that the ill-defined term $\Delta \mathcal{S}$ occurs at the first order in $\hbar$, so it can be ignored as a first approximation. That is, we are only going to assume that $\mathcal{S}$ satisfies the CME $(\mathcal{S}, \mathcal{S}) = 0$. The functional integral is then defined in terms of Feynman diagrams, and, a posteriori, we check that it is indeed invariant under deformations of the gauge-fixing Lagrangian. (More correctly, one should regularize the theory while defining the Feynman diagrams and consistently check the QME in the regularized version; see [54, 55] and especially [33, 34]).

With this caveat, we can now make sense of the partition function $Z := \int_{\mathcal{L}_t} e^{\frac{i}{\hbar}S}\rho$, where $\rho$ is the reference half-density. The next goal is to consider functionals $\mathcal{O}$ on $\mathcal{F}$ for which the expectation value

$$\langle \mathcal{O} \rangle := \frac{1}{Z} \int_{\mathcal{L}_t} e^{\frac{i}{\hbar}S} \mathcal{O} \rho$$

is also invariant under deformations of the gauge-fixing Lagrangian. By the BV property (BV2.2), this happens if $\Delta\left(e^{\frac{i}{\hbar}\mathcal{S}}\mathcal{O}\right) = 0$, that is, assuming the QME, if

$$Q\mathcal{O} - i\hbar \Delta \mathcal{O} = 0,$$

---

[7]In physics, one often realizes $\mathcal{L}_t$ as the graph in $T^*[-1]\mathcal{M}$ of an odd function $\Psi_t$ of ghost number $-1$—the gauge-fixing fermion.

where, as above, $Q = -(\mathcal{S}, \ )$. We call such a functional a quantum BV observable.

*Remark* 3.2 (Quantum BV cohomology). It is convenient to introduce the operator $\Omega := Q - i\hbar\Delta$, which is a coboundary operator thanks to the QME. BV observables are by definition $\Omega$-closed. On the other hand, by property (BV2.1), the expectation value of an $\Omega$-exact functional vanishes. As a consequence, a BV theory, with a choice of a lagrangian submanifold up to deformations, is a linear functional on the cohomology $H_\Omega$ of $\Omega$ (to be precise, we should only consider classes with a representative for which the functional integral makes sense.).

Again, the ill-defined term $\Delta \mathcal{O}$ occurs at the first order in $\hbar$, so we will at first ignore it. We therefore define a classical BV observable as a functional $\mathcal{O}$ on $\mathcal{F}$ satisfying $Q\mathcal{O} = 0$.

It follows that $\mathcal{O}|_{\mathcal{L}_0}$ is a functional on $F$ that is invariant under the infinitesimal symmetries. Usually one has to work the other way around, starting with a classical invariant functional and looking for an extension to a classical BV observable. Our goal will be to construct a classical BV observable $\mathcal{O}_\sigma$ for $BF$ theory with cosmological term which, on $\mathcal{L}_0$, reduces to (2.5).

3.4. **The BV pushforward.** In this section we describe a partial BV integration (BV pushforward) which is needed for the rest of the paper; see [31, 51, 54, 55, 33, 26, 34, 27].

We assume that our odd symplectic manifold $(\mathcal{F}, \omega)$ is the product[8] of two odd symplectic manifolds $(\mathcal{F}_1, \omega_1)$ and $(\mathcal{F}_2, \omega_2)$: $\mathcal{F} = \mathcal{F}_1 \times \mathcal{F}_2$, $\omega = \pi_1^*\omega_1 + \pi_2^*\omega_2$, with $\pi_i$ the canonical projection $\mathcal{F} \to \mathcal{F}_i$. The canonical BV Laplacian $\triangle$ is then also a sum $\triangle = \triangle_1 + \triangle_2$. If $\rho$ is a half-density on $\mathcal{F}$ and $\mathcal{L}_2$ is a lagrangian submanifold of $\mathcal{F}_2$, then $\int_{\mathcal{L}_2} \rho$ is a half-density on $\mathcal{F}_1$ (assuming that the integral converges). Moreover, we have:

**BVpf1:** $\triangle_1 \int_{\mathcal{L}_2} \rho = \int_{\mathcal{L}_2} \triangle \rho$, and

**BVpf2:** $\frac{d}{dt} \int_{\mathcal{L}_{2,t}} \rho$ is $\triangle_1$-exact if $\rho$ is $\triangle$-closed.

This means that the partial BV integration is a well-defined[9] operation in the cohomologies of the BV Laplacians (as long as we choose a deformation class of lagrangian submanifolds of $\mathcal{F}_2$). This operation is called the BV pushforward.

We now fix a reference, $\triangle$-closed, nowhere vanishing half-density $\rho$ on $\mathcal{F}$. It follows that $\rho_1 := \int_{\mathcal{L}_2} \rho$ is also $\triangle$-closed and nowhere vanishing.

---

[8]In [27] the case of nontrivial fibrations is also discussed.

[9]To be precise, it is well-defined assuming that the integral converges.

If $\mathcal{S}$ is a solution of the QME master equation, we define $\mathcal{S}_1$ via

$$e^{\frac{i}{\hbar}\mathcal{S}_1}\rho_1 = \int_{\mathcal{L}_2} e^{\frac{i}{\hbar}\mathcal{S}}\rho.$$

The new BV action $\mathcal{S}_1$ (called the effective action) solves the QME on $\mathcal{F}_1$ and is defined up to quantum BV transformations (i.e., $e^{\frac{i}{\hbar}\mathcal{S}_1}$ is defined up to $\Delta_1$-exact terms). We call $(\mathcal{F}, \omega, \mathcal{S})$ the parent theory and $(\mathcal{F}_1, \omega_1, \mathcal{S}_1)$ the effective theory (we borrow the terminology from [11]).

*Remark* 3.3 (Pushforward of observables). If $\mathcal{O}$ is a quantum BV observable for the parent theory (i.e., $\Delta\left(e^{\frac{i}{\hbar}\mathcal{S}}\mathcal{O}\right) = 0$), we define the corresponding quantum BV observable $\mathcal{O}_1$ for the effective theory via

$$(3.3) \qquad e^{\frac{i}{\hbar}\mathcal{S}_1}\mathcal{O}_1\rho_1 = \int_{\mathcal{L}_2} e^{\frac{i}{\hbar}\mathcal{S}}\mathcal{O}\rho.$$

This way, one can construct quantum BV observables for the the effective theory starting from quantum BV observables for the parent theory. We will use this procedure in Section 7.

3.4.1. *BV quasiisomorphism.* More generally, we can think of (3.3) as a map from functions on $\mathcal{F}$ to functions on $\mathcal{F}_1$. By property (BVpf1), we see that this is actually a chain map with respect to the coboundary operators $\Omega = Q + i\hbar\Delta$ and $\Omega_1 Q_1 + i\hbar\Delta_1$, see Remark 3.2. As such, it induces a map in the corresponding cohomologies $H_\Omega$ and $H_{\Omega_1}$.

If this induced map is an isomorphism, one says that the chain map between functionals is a quasiisomorphism. This is the case under some assumptions which are (formally) satisfied by all examples in this note. One simple setting is when $\mathcal{S}$ is a perturbation of abelian $BF$ theory and the gauge-fixing Lagrangian comes from a chain homotopy operator. Strictly speaking, this is a theorem [55, 54] (see also [27] for a more general discussion) in the finite-dimensional case—in the infinite-dimensional one, the BV Laplacian is not even properly defined—but we formally extend it to field theory. Note that the semi-classical part of this (i.e., the restriction to only trees in the expansion in Feynman diagrams) yields back the homotopy transfer by Merkulov [53] and Kontsevich–Soibelman [48].

3.5. **Construction of observables.** Another way to obtain quantum BV observables for a BV theory is the following [31]. Suppose that $\mathcal{S} = \pi_1^*\mathcal{S}_{(1)} + \mathcal{S}_{(2)}$ with $\mathcal{S}_{(1)}$ solving the QME on $\mathcal{F}_1$ (we still also assume that $\mathcal{S}$ solves the QME on $\mathcal{F}$). We also assume that $\rho = \rho_1\rho_2$, with $\rho_1$ the induced reference half-density on $\mathcal{F}_1$ and $\rho_2$ a reference, $\triangle$-closed, nowhere vanishing half-density on $\mathcal{F}_2$. Then

$$e^{\frac{i}{\hbar}\mathcal{S}_1}\rho_1 = e^{\frac{i}{\hbar}\mathcal{S}_{(1)}}\rho_1 \int_{\mathcal{L}_2} e^{\frac{i}{\hbar}\mathcal{S}_{(2)}}\rho_2,$$

and we get the quantum BV observable

$$(3.4) \qquad \mathcal{O}_1 = \int_{\mathcal{L}_2} e^{\frac{i}{\hbar}\mathcal{S}_{(2)}}\rho_2$$

for the BV theory $(\mathcal{F}_1, \omega_1, \mathcal{S}_{(1)})$ (note that this is not the effective theory). We will use this procedure in Section 4.

3.6. **Residual fields.** A special case of the BV pushforward construction is when $\mathcal{F}_1$ is finite-dimensional. In this case, following [27], we call it the space of residual fields.

There can be different choices, also with different dimensions, of spaces of residual fields. Assuming they are all quasiisomorphic to each other, we consider them as fully equivalent. Usually there is a minimal choice, but in some cases (e.g., when one wants to glue theories along a boundary [27]) it is not the most convenient.

More generally, we think of $\mathcal{F}_2$ as a product $\mathcal{F}_{\text{res}} \times \widehat{\mathcal{F}}_2$, where $\mathcal{F}_{\text{res}}$ is finite-dimensional. In this case, when we speak of a BV pushforward along $\mathcal{F}_2$, we actually mean a BV pushforward along $\widehat{\mathcal{F}}_2$ yielding an effective theory on $\mathcal{F}_1 \times \mathcal{F}_{\text{res}}$.

In particular, the construction of Section 3.5 yields a quantum BV observable $\mathcal{O}_1$, see equation (3.4), for the BV theory $(\mathcal{F}_1, \omega_1, \mathcal{S}_{(1)})$ parametrized over $(\mathcal{F}_{\text{res}}, \omega_{\text{res}}, \mathcal{S}_{\text{res}})$. An appropriate integration over the latter space may sometimes be performed (one notable example is described in [14]).

## 4. BV SURFACE OBSERVABLES FOR $BF$ THEORY

$BF$ theory is an example of an AKSZ [1] theory, so its BV version is easily constructed. We give the final answer, as it is immediate to verify that the BV action we write extends the classical action (2.1) and satisfies the master equation. We first introduce the "superfields"

$$(4.1) \qquad \begin{aligned} \mathcal{A} &= c + A + B^+ + \tau^+ + \phi^+, \\ \mathcal{B} &= \phi + \tau + B + A^+ + c^+. \end{aligned}$$

In each superfield, the fields (with the exception of $A$, which is a connection 1-form) are ad-valued differential forms ordered by increasing form degree (with $c$ and $\phi$ 0-forms and with $\phi^+$ and $c^+$ 4-forms). They are assigned ghost numbers in decreasing order starting with $c$ of ghost number 1 and $\phi$ of ghost number 2 (in other words, a summand in $\mathcal{A}$ has ghost number equal to 1 minus its form degree, and a summand in $\mathcal{B}$ has ghost number equal to 2 minus its form degree). Here $A$ and $B$ are the classical fields already appearing in (2.1), whereas $c$ and $\tau$ are the ghosts corresponding to the infinitesimal symmetries (2.3). We have to introduce the ghost-for-ghost $\phi$ in order to have the infinitesimal action injective also on-shell. The fields with a + superscripts are the momenta. More precisely, we have the BV symplectic form

$$\omega := \int_M \delta\mathcal{B}\, \delta\mathcal{A},$$

where $\int_M$ selects the summands of form degree equal to 4 (therefore, of ghost number $-1$). The BV action is

$$(4.2) \qquad \mathcal{S} := \int_M \left( \mathcal{B} F_\mathcal{A} - \frac{\lambda}{2} \mathcal{B}\mathcal{B} \right),$$

where we use the same convention as for the BV symplectic form and $F_\mathcal{A} := \mathrm{d}\mathcal{A} + \frac{1}{2}[\mathcal{A}, \mathcal{A}]$ is the "curvature of the superconnection $\mathcal{A}$."

The hamiltonian vector field $Q$ of $\mathcal{S}$, see (3.1), can be easily computed. We should only be careful with signs.

*Remark* 4.1 (Sign conventions). Throughout the paper, we use the total degree, i.e., the sum of form and ghost degree, to determine the parity of an object and the consequent sign rules. According to this convention, each summand in $\mathcal{A}$ is odd and each summand in $\mathcal{B}$ is even. Moreover, the operators $\mathrm{d}_A$ and $Q$ are odd, but the contraction $\iota_Q$ is even. Finally, we need the odd operator $\mathrm{d}_\mathcal{A}$, which is locally defined as $\mathrm{d} + [\mathcal{A},\ ]$.

Following these conventions, we have[10]

$$(4.3\mathrm{a}) \qquad Q\mathcal{A} = F_\mathcal{A} - \lambda\mathcal{B},$$
$$(4.3\mathrm{b}) \qquad Q\mathcal{B} = \mathrm{d}_\mathcal{A}\mathcal{B}.$$

One can easily see that $(\mathcal{S}, \mathcal{S}) = -Q\mathcal{S} = 0$. Moreover, formally we also have $\Delta\mathcal{S} = 0$ if we assume the Lie algebra to be unimodular.

---

[10]In particular, we get

$$QA = \mathrm{d}_A c - \lambda\tau,$$
$$QB = \mathrm{d}_A \tau + [c, B] + [B^+, \phi].$$

As a result of our sign conventions, we then have, e.g., $QF_A = -\mathrm{d}_A \delta A = -[F_A, c] + \lambda\, \mathrm{d}_A \tau$. Consistency with (2.3) is recovered by setting $B^+$ to zero and using the relations

$$Q = \epsilon\delta, \quad c = \epsilon\gamma, \quad \tau = \epsilon\theta,$$

with $\epsilon$ an odd variable.

We denote by $\mathcal{F}_M$ the BV space of fields. If $A_0$ is a reference connection, we have $\mathcal{F}_M = A_0 + \Omega^\bullet(M, \mathrm{ad}P)[1] \oplus \Omega^\bullet(M, \mathrm{ad}P)[2]$.[11]

To get the BV action corresponding to $J$ in (2.4), we follow a similar strategy as above. First we introduce the superfields
$$a = \psi + \alpha + \beta^+,$$
$$b = \beta + \alpha^+ + \psi^+,$$

following the same conventions for form degree and ghost number (just note that $\psi$ has ghost number 1, whereas $\beta$ has ghost number 0). All the summands here are sections of the ad-bundle pulled back to $\Sigma$. We denote this space of fields by $\mathcal{F}_\Sigma = \Omega^\bullet(\Sigma, \mathrm{ad}P)[1] \oplus \Omega^\bullet(\Sigma, \mathrm{ad}P)$.

We extend the BV symplectic form $\omega$ by adding a term $\omega_\Sigma := \int_\Sigma \delta b\, \delta a$. Finally, we consider the BV action
$$\mathcal{J}[a, b; \mathcal{A}, \mathcal{B}] := \int_\Sigma \left( b\, \mathrm{d}_\mathcal{A} a + \frac{\lambda}{2} b[a, a] + b\mathcal{B} \right)$$

and define
$$\widehat{\mathcal{S}} := \mathcal{S} + \mathcal{J}.$$

We denote by $\widehat{Q}$ its hamiltonian vector field.

**Lemma 4.2.** $\widehat{\mathcal{S}}$ *satisfies the classical master equation* $(\widehat{\mathcal{S}}, \widehat{\mathcal{S}}) = -\widehat{Q}\widehat{\mathcal{S}} = 0$.

*Proof.* We denote by $R$ the hamiltonian vector field of $\mathcal{J}$. An easy computation yields
$$Ra = \mathrm{d}_\mathcal{A} a + \frac{\lambda}{2}[a, a] + \mathcal{B},$$
$$Rb = \mathrm{d}_\mathcal{A} b + \lambda[a, b],$$
$$R\mathcal{A} = b\, \delta_\Sigma,$$
$$R\mathcal{B} = [a, b]\, \delta_\Sigma,$$

where $\delta_\Sigma$ is the delta 2-form localized on $\Sigma$. We have $\widehat{Q} = Q + R$ and
$$\widehat{Q}\widehat{\mathcal{S}} = (Q + R)(\mathcal{S} + \mathcal{J}) = Q\mathcal{S} + Q\mathcal{J} + R\mathcal{S} + R\mathcal{J}.$$

We already mentioned that $Q\mathcal{S} = 0$. An explicit computation shows
$$Q\mathcal{J} = R\mathcal{S} = \int_\Sigma (b[F_\mathcal{A}, a] - \lambda b[\mathcal{B}, a] + b\, \mathrm{d}_\mathcal{A} \mathcal{B}).$$

Finally, another explicit computation yields $R\mathcal{J} = -2Q\mathcal{J}$. □

---

[11]In particular, if $P$ is trivial, we may choose $A_0$ as the zero connection and get $\mathcal{F}_M = (\Omega^\bullet(M)[1] \oplus \Omega^\bullet(M)[2]) \otimes \mathfrak{g}$.

The above proof is a bit unsatisfactory, as some terms are actually singular. In fact, in the computation of $R\mathcal{S}$, we also get the terms

$$\int_\Sigma (b[b,a] + b[a,b])\delta_\Sigma.$$

The delta 2-form $\delta_\Sigma$ evaluated on $\Sigma$ is singular. It should be regularized as a Poincaré dual of $\Sigma$. On the other hand, the term in the round bracket clearly vanishes, so formally this term is zero.

Remaining at the same heuristic level, we can however also conclude that $\Delta\mathcal{J} = 0$ under the assumptions that the Lie algebra is unimodular and that $\sigma\colon \Sigma \to M$ is an embedding. This also implies that $\Delta\widehat{\mathcal{S}} = 0$, so $\widehat{\mathcal{S}}$ satisfies the quantum master equation.

As a consequence, a BV pushforward on the fields $a$ and $b$ as in (3.4) formally yields a quantum BV observable $\mathcal{O}_\sigma$ for $BF$ theory, replacing the naive definition (2.5). More precisely,

$$\mathcal{O}_\sigma := \int_{\mathcal{L}_2} e^{\frac{i}{\hbar}\mathcal{J}}\rho_2.$$

We will avoid the issues of the heuristic computations arising in this section by using the perturbative formulation of the observable $\mathcal{O}_\sigma$ in Section 5.

4.1. **Isotopy invariance.** We conclude this section with a heuristic proof that the expectation value of $\mathcal{O}_\sigma$ is a knot invariant. This has to be expected as we are dealing with a topological field theory. It is however instructive to view a direct argument using the BV formalism.

Let $\phi\colon M \times [0,1] \to M$ be an isotopy relating two embeddings $\sigma_0$ and $\sigma_1$ of $\Sigma$ into $M$. Let $\xi_t$ be the corresponding vector field at $t \in [0,1]$.

In case the principal bundle $P$ is trivial and we regard all the fields, including the connection $A$, as $\mathfrak{g}$-valued differential forms (i.e., we work around the trivial connection), we define

$$\mathcal{U}_t := \int_\Sigma \left(b\left[\iota_{\xi_t}\mathcal{A}, a\right] + b\,\iota_{\xi_t}\mathcal{B}\right).$$

A simple computation, by the same heuristics as before, shows that

$$\widehat{Q}\mathcal{U}_t = \int_\Sigma \left(b\left[\mathsf{L}_{\xi_t}\mathcal{A}, a\right] + b\,\mathsf{L}_{\xi_t}\mathcal{B}\right) = \frac{\mathrm{d}}{\mathrm{d}t}\mathcal{J}_t,$$

where $\mathsf{L}$ denotes the Lie derivative. Since, also formally, $\Delta\mathcal{U}_t = 0$, we get

$$\frac{\mathrm{d}}{\mathrm{d}t}\mathcal{O}_\sigma = \Omega\int_{\mathcal{L}_2} e^{\frac{i}{\hbar}\mathcal{J}}\mathcal{U}_t\rho_2,$$

so the the *t*-derivative of the expectation value of $\mathcal{O}_\sigma$ in $BF$ theory vanishes.

If the principal bundle $P$ is not trivial, we lift the vector fields $\xi_t$ to $P$ by a reference connection $A_0$. In this case, we have

$$\frac{\mathrm{d}}{\mathrm{d}t}\mathcal{J}_t = \int_\Sigma \left(b\,[\iota_\xi F_{A_0} + \mathsf{L}_{\xi_t}^{A_0}\widehat{\mathcal{A}}, a] + b\,\mathsf{L}_{\xi_t}^{A_0}\mathcal{B}\right),$$

where $\widehat{\mathcal{A}} = \mathcal{A} - A_0$ and $\mathsf{L}_{\xi_t}^{A_0} = [\iota_\xi, \mathrm{d}_{A_0}]$. We now define

$$\mathcal{U}_t := \int_\Sigma \left(b\,[\iota_{\xi_t}\widehat{\mathcal{A}}, a] + b\,\iota_{\xi_t}\mathcal{B}\right)$$

and verify that $\widehat{Q}\mathcal{U}_t = \frac{\mathrm{d}}{\mathrm{d}t}\mathcal{J}_t$. The rest of the argument is as in the trivial case.

4.2. **Topological deformations and other observables.** The setting presented so far can be modified to a certain extent. This will be useful in connection with Yang–Mills theory (see Section 7).

4.2.1. *Topological deformations.* The functional

$$\mathcal{S}_{\text{top}}^\kappa := \frac{\kappa}{2}\int_M F_\mathcal{A} F_\mathcal{A},$$

where $\kappa$ is a parameter, satisfies $\delta\mathcal{S}_{\text{top}}^\kappa = 0$ (for $M$ closed), so it can be used to deform the $BF$ action (4.2) to

$$\mathcal{S}^\kappa := \mathcal{S} + \mathcal{S}_{\text{top}}^\kappa,$$

which still satisfies the classical (and formally the quantum) master equation. Also note that the hamiltonian vector field of $\mathcal{S}^\kappa$ is the same as that of $\mathcal{S}$. As a consequence,

$$\widehat{\mathcal{S}}^\kappa := \mathcal{S}^\kappa + \mathcal{J}$$

still satisfies the classical (and formally the quantum) master equation for any $\kappa$. Therefore, the BV pushforward of $\mathrm{e}^{\frac{\mathrm{i}}{\hbar}\mathcal{J}}$ defines a surface observable for $\mathcal{S}^\kappa$ with any $\kappa$.

4.2.2. *Other surface observables.* Consider the transformation

$$\mathcal{A} \mapsto \mathcal{A}, \qquad \mathcal{B} \mapsto \mathcal{B} + \gamma F_\mathcal{A},$$

for some parameter $\gamma$. This is a quantum BV transformation (i.e., a symplectomorphisms that, formally, does not change the reference "Lebesgue" half-density). The action is transformed accordingly, and a simple computation yields

$$\mathcal{S}^\kappa \to (1-\lambda\gamma)\mathcal{S}' + \mathcal{S}_{\text{top}}^{\kappa+\gamma-\frac{\lambda\gamma^2}{2}}$$

with
$$\mathcal{S}' := \int_M \left( \mathcal{B} F_{\mathcal{A}} - \frac{\lambda'}{2} \mathcal{B}\mathcal{B} \right)$$
and
$$\lambda' = \frac{\lambda}{1 - \gamma\lambda}.$$

It follows that $(1 - \lambda\gamma)(\mathcal{S}' + \mathcal{J}')$ and $(1 - \lambda\gamma)(\mathcal{S}' + \mathcal{J}') + \mathcal{S}_{\text{top}}^{\kappa + \gamma - \frac{\lambda\gamma^2}{2}}$, with $\mathcal{J}'$ obtained from $\mathcal{J}$ replacing $\lambda$ with $\lambda'$, still satisfy the classical (and formally the quantum) master equation.

We can now revert the transformation getting back $\mathcal{S}^\kappa$. This transforms $(1 - \lambda\gamma)\mathcal{J}'$ into

$$(4.4) \qquad \mathcal{J}_\gamma := (1 - \lambda\gamma) \int_\Sigma \left( b\, \mathrm{d}_{\mathcal{A}} a + \frac{\lambda'}{2} b[a, a] + b\mathcal{B} - \gamma b F_{\mathcal{A}} \right).$$

In conclusion,
$$\widehat{\mathcal{S}}_\gamma^\kappa := \mathcal{S}^\kappa + \mathcal{J}_\gamma$$
still satisfies the classical (and formally the quantum) master equation for any $\kappa$ and $\gamma$. Therefore, for every $\gamma$, the BV pushforward of $e^{\frac{i}{\hbar} \mathcal{J}_\gamma}$ defines a surface observable for $\mathcal{S}^\kappa$ with any $\kappa$.

## 5. Perturbative expansion

To start the perturbative expansion, we have to fix a critical point $(A_0, B_0)$ for the ambient theory and then pick a critical point $(\alpha_0, \beta_0)$ for the theory on $\Sigma$ evaluated at $(A_0, B_0)$.[12]

We prefer however to consider also $\lambda$ as a perturbative parameter. Therefore, we assume $F_{A_0} = 0$ and $\mathrm{d}_{A_0} B_0 = 0$. We can then rewrite the action on $\Sigma$ at $(A_0, B_0)$ as

$$\mathcal{J}[a, b; A_0, B_0] := \int_\Sigma (b\, \mathrm{d}_{A_0} a + b B_0) + O(\lambda).$$

A critical point for the classical fields of the theory on $\Sigma$ is now a solution of
$$\mathrm{d}_{A_0} \beta_0 = 0 \quad \text{and} \quad \mathrm{d}_{A_0} \alpha_0 + B_0 = 0,$$

---

[12]In principle, a critical point for the theory is a quadruple $(A_0, B_0, \alpha_0, \beta_0)$ satisfied the coupled equations; yet we want to think of $\mathcal{J}$ as defining an observable for the ambient theory, so we first look for a critical point $(A_0, B_0)$ for the ambient theory that ignores the singularity on $\Sigma$ due to $\mathcal{J}$ and then a related critical point $(\alpha_0, \beta_0)$.

and we complete it to a critical point for the BV theory by setting all the other BV fields to zero. If we write

(5.1)
$$\begin{aligned} a &= \alpha_0 + \widehat{a}, \\ b &= \beta_0 + \widehat{b}, \\ \mathcal{A} &= A_0 + \widehat{\mathcal{A}}, \\ \mathcal{B} &= B_0 + \widehat{\mathcal{B}}, \end{aligned}$$

where the hatted fields are the perturbations, we get

$$\mathcal{J}[a,b;\mathcal{A},\mathcal{B}] = \int_\Sigma \widehat{b}\, \mathrm{d}_{A_0} \widehat{a} + \text{perturbations.}$$

The gauge-fixing Lagrangian $\mathcal{L}_2$, and the choice of residual fields, have then to be chosen in such way that on $\mathcal{L}_2$ the operator $\mathrm{d}_{A_0}$ can be inverted.

*Remark* 5.1 (Result of the perturbative expansion). The gauge-fixed BV integral will ultimately yield a "function" $\mathcal{O}_\sigma$ of the background fields $(\alpha_0, \beta_0)$ and of the ambient fields $(\mathcal{A}, \mathcal{B})$. To be more precise, we will have to split the hatted fields $(\widehat{a}, \widehat{b})$ into residual fields $(\underline{a}, \underline{b})$ and fluctuations $(\widetilde{a}, \widetilde{b})$,

(5.2) $$\widehat{a} = \underline{a} + \widetilde{a}, \qquad \widehat{b} = \underline{b} + \widetilde{b},$$

where the residual fields parametrize the cohomology of $\mathrm{d}_{A_0}$. The result of the gauge-fixed BV integral over the fluctuations $(\widetilde{a}, \widetilde{b})$ will then yield $\mathcal{O}_\sigma$ as a half-density on the space of residual fields $(\underline{a}, \underline{b})$, parametrized by the background fields $(\alpha_0, \beta_0)$ and the ambient fields $(\mathcal{A}, \mathcal{B})$. The starting point of the perturbative computation is the computation of the Berezinian of the quadratic form: this yields the analytic torsion for the complex with differential $\mathrm{d}_{A_0}$, which is indeed a half-density over the space of residual fields $(\underline{a}, \underline{b})$. The full perturbative expansion multiplies this torsion by an expansion in Feynman diagrams.

5.1. **The perturbative expansion in the trivial case.** For simplicity, we will only consider the case when the principal bundle $P$ is trivial and will work around $A_0 = 0$ and $B_0 = 0$. In this case $\mathrm{d}_{A_0}$ is just the exterior derivative $\mathrm{d}$ (tensor the identity on $\mathfrak{g}$). We fix a Riemannian metric on $\Sigma$ and consider the related Hodge decomposition

$$\Omega^\bullet(\Sigma) = \mathrm{Harm}^\bullet(\Sigma) \oplus \mathrm{d}\Omega^{\bullet-1}(\Sigma) \oplus \mathrm{d}^*\Omega^{\bullet+1}(\Sigma),$$

where $\mathrm{d}^*$ is the Hodge codifferential and Harm denotes the space of harmonic forms. According to Section 3.6, we rewrite the BV space of fields $\mathcal{F}_\Sigma = (\Omega^\bullet(\Sigma)[1] \oplus \Omega^\bullet(\Sigma)) \otimes \mathfrak{g}$ as

$$\mathcal{F}_\Sigma = \mathcal{F}_{\Sigma,\text{res}} \times \widehat{\mathcal{F}}_\Sigma$$

with $\mathcal{F}_{\Sigma,\text{res}} = (\text{Harm}^\bullet(\Sigma)[1] \oplus \text{Harm}^\bullet(\Sigma)) \otimes \mathfrak{g}$ and $\widehat{\mathcal{F}}_\Sigma$ the space of exact and coexact $\mathfrak{g}$-valued forms. Inside $\widehat{\mathcal{F}}_\Sigma$ we choose the Lagrangian submanifold

$$\mathcal{L}_2 \coloneqq (\text{im }\mathrm{d}^*) \otimes \mathfrak{g}.$$

5.1.1. *The propagator.* To construct the propagator, we focus first on the abelian theory and rewrite $\int_\Sigma \widehat{b}\,\mathrm{d}\widehat{a}$ as $\langle \widehat{b}, *^{-1}\mathrm{d}\widehat{a} \rangle$, where $\langle \eta, \zeta \rangle = \int_\Sigma \eta * \zeta$ is the Hodge pairing of forms. The propagator is then the inverse of $*^{-1}\mathrm{d}$ on $\mathrm{d}^*\Omega^\bullet(\Sigma)$. We look for the inverse $L$ of $*\mathrm{d}$ which is related to the propagator by a sign operator. We claim that

$$L = *\mathrm{d}G,$$

where $G$ is the inverse of the Laplace operator $\mathrm{d}^*\mathrm{d} + \mathrm{d}\mathrm{d}^*$, which is invertible in the orthogonal complement of $\text{Harm}^\bullet(\Sigma)$. In fact, we have $*\mathrm{d}L = \mathrm{d}^*\mathrm{d}G = G\mathrm{d}^*\mathrm{d}$. Therefore, for any $\gamma$,

$$*\mathrm{d}L\mathrm{d}^*\gamma = G\mathrm{d}^*\mathrm{d}\mathrm{d}^*\gamma = G(\mathrm{d}^*\mathrm{d} + \mathrm{d}\mathrm{d}^*)\mathrm{d}^*\gamma = \mathrm{d}^*\gamma,$$

so $*\mathrm{d}L$ acts as the identity on the image of $\mathrm{d}^*$. This is the propagator as considered, in Chern–Simons theory, by Axelrod and Singer [6, 7].

It is convenient to extend $G$ to the de Rham complex $\Omega^\bullet(\Sigma)$ defining it to be zero on $\text{Harm}^\bullet(\Sigma)$. Next we define

(5.3) $$K \coloneqq \mathrm{d}^*G.$$

It is clear that $L$ is, up to a sign operator, the restriction to $\mathrm{d}^*\Omega^\bullet(\Sigma)$ of $K*$. Moreover, we have

(5.4) $$\mathrm{d}K + K\mathrm{d} = \text{id} - \pi_{\text{Harm}},$$

where $\pi_{\text{Harm}}$ is the orthogonal projection to $\text{Harm}^\bullet(\Sigma)$. In other words, $K$ is a parametrix for $\mathrm{d}$. Using another terminology, $(K, \iota, \pi_{\text{Harm}})$, where $\iota$ is the inclusion map of $\text{Harm}^\bullet(\Sigma)$, is a deformation retract for the de Rham complex $(\Omega^\bullet(\Sigma), \mathrm{d})$. This is at the basis of the homotopy transfer, which will correspond to the tree-level Feynman diagrams in the perturbative expansion.

5.1.2. *The soft propagator.* A more general and flexible definition of propagator, which we will call a soft propagator, is any operator $K$ satisfying (5.4).[13] A propagator in this sense can always be constructed

---

[13]Axelrod and Singer's propagator has further nice properties, sometimes called the side conditions of an SDR (strong/special deformation retract):

$$K^2 = 0, \qquad K\iota = 0, \qquad \pi_{\text{Harm}}K = 0.$$

by soft methods as in [17] (see also [19]). Namely, one can write

$$K\alpha = -\pi_{2*}(\eta\,\pi_1^*\alpha), \qquad \alpha \in \Omega^\bullet(\Sigma),$$

where $\pi_1$ and $\pi_2$ are the two projections $\Sigma \times \Sigma \to \Sigma$ and the lower $*$ denotes integration along the fiber (pushforward). Here $\eta$ is the (soft) propagator kernel as a 1-form[14] on $\Sigma \times \Sigma$ with a singularity on the diagonal $\triangle_\Sigma$. It is then convenient to remove the diagonal and to consider the configuration space $C_2^0(\Sigma) := \Sigma \times \Sigma \setminus \triangle_\Sigma$, on which $\eta$ is smooth.

Even better, we consider the compactification $C_2(\Sigma)$ obtained as the closure of $C_2^0(\Sigma)$ inside $\Sigma \times \Sigma \times \mathrm{Bl}(\triangle_\Sigma)$, where $\mathrm{Bl}(\triangle_\Sigma)$ is the differential-geometric blowup obtained by replacing $\triangle_\Sigma$ inside $\Sigma \times \Sigma$ with its unit normal bundle.[15] We keep denoting by $\pi_1$ and $\pi_2$ the two projections to $\Sigma$. Now, $\pi_2^{-1}(x)$ is a compact manifold for every $x$ in $\Sigma$, so the the pushforward $\pi_{2*}$ is integration along a compact manifold. This ensures both convergence (assuming of course $\eta$ smooth) and the possibility of exchanging $\pi_{2*}$ and derivatives along $\Sigma$. Up to a sign, we then have, by Stokes' theorem,

$$\mathrm{d}_\Sigma \pi_{2*} - \pi_{2*} \mathrm{d}_{C_2(\Sigma)} = \pi_*^\partial,$$

where $\pi^\partial$ is the projection $\partial C_2(\Sigma) \to \Sigma$. The boundary $\partial C_2(\Sigma)$ can be identified with the unit tangent bundle of $\Sigma$.

The analogue of having harmonic forms defined by the metric is now a choice of basis $([\chi_i])_i$ for $H^\bullet(\Sigma)$ with a choice of representatives $\chi_i \in \Omega^\bullet(\Sigma)$. We denote by

$$\sigma_{ij} := \int_\Sigma \chi_i\,\chi_j$$

the entries of the nondegenerate pairing on $H^\bullet(\Sigma)$ in the basis $([\chi_i])_i$. We will denote by $\sigma^{ij}$ the entries of its inverse.

The soft propagator $\eta$ is then defined as a smooth 1-form on $C_2(\Sigma)$ satisfying the following properties:

---

We will not demand them in general, even though they may be useful for more explicit computations.

[14]In general, for $\Sigma$ of dimension $k$, $\eta$ is a $(k-1)$-form.

[15]In practice, $C_2(\Sigma)$ is easily described in terms of coordinates: the usual ones from $\Sigma \times \Sigma$ away from the diagonal, whereas near the diagonal we take polar coordinates $(r, \theta)$ for one of the two points in $\Sigma$ with respect to the other and compactify by allowing $r$ also to attain the value zero.

(1) $\mathrm{d}\eta = \pi^*\chi$, where $\pi$ is the projection $C_2(\Sigma) \to \Sigma \times \Sigma$, and $\chi$ is the representative of the Euler class of $\Sigma$ given by
$$\chi = -\sum_{ij} \pi_1^*\chi_i \, \pi_2^*\chi_j \, \sigma^{ij}$$
with $\pi_1$ and $\pi_2$ the projections $\Sigma \times \Sigma \to \Sigma$.
(2) The restriction of $\eta$ to $\partial C_2(\Sigma)$ is a global angular form. In particular, $\int_{\partial C_2(\Sigma)} \eta = 1$.

It is shown in [17] (see also [19] for a review) how to construct such an $\eta$. Moreover, one can always ensure that $T^*\eta = \eta$, where $T$ is the map that exchanges the two points.

*Remark* 5.2 ([6, 7]). The Axelrod and Singer's propagator $K$, see (5.3), also has a kernel $\eta$ satisfying these properties.

5.1.3. *Feynman diagrams.* The (soft) propagator kernel $\eta$ can also be identified with the 2-point expectation value (in the theory defined by $\mathcal{J}$) of the fluctuations $\widehat{a}$ and $\widehat{b}$ and therefore with the edges in the Feyman-diagram expansion. More precisely, accordingly to (5.2), we write
$$\widehat{a} = \underline{a} + \widetilde{a}, \qquad \widehat{b} = \underline{b} + \widetilde{b},$$
with $\underline{a} = \sum_i z^i \chi_i$ and $\underline{b} = \sum_i w^i \chi_i$, where $(z^i, w^i)$ are coordinates on the finite-dimensional BV manifold $H^\bullet(\Sigma)[1]$ with respect to the basis $([\chi_i])_i$. The BV form is just $\omega_{H^\bullet(\Sigma)[1]} = \sum_{ij} \mathrm{d}z^i \, \mathrm{d}w^j \, \sigma_{ij}$. Since the kinetic term $\int_\Sigma \widehat{b} \, \mathrm{d}_{A_0} \widehat{a}$ is just $\int_\Sigma \widetilde{b} \, \mathrm{d}\widetilde{a}$, we now have
$$\eta = \langle \pi_1^*\widetilde{a} \, \pi_2^*\widetilde{b} \rangle$$
with $\pi_1$ and $\pi_2$ the projections $C_2(\Sigma) \to \Sigma$.

The Feynman diagrams of the theory are now obtained, as usual, by recognizing in the expansion of $\mathcal{J}[a, b; \mathcal{A}, \mathcal{B}]$ the vertices in $\widetilde{a}$ and $\widetilde{b}$, depending parametrically on $(\underline{a}, \underline{b}, \mathcal{A}, \mathcal{B})$. To each vertex we then associate a point on $\Sigma$, to be integrated, and to each possible pairing of an $\widetilde{a}$ with a $\widetilde{b}$ we associate a propagator $\eta$. Naively each Feynman diagram yields an integral over a product of $n$ copies of $\Sigma$, where $n$ is the number of vertices. The only mild regularization needed here consists in restricting the integration to the open coniguration space $C_n^0(\Sigma)$ obtained by removing from $\Sigma^n$ all diagonals.

The crucial point is that there exists a nice compactification $C_n(\Sigma)$ of $C_n^0(\Sigma)$, due to Axelrod and Singer [6, 7], elaborating on [40], (see also [18] and [66]), with the following properties:

(1) Each $C_n(\Sigma)$ is a smooth compact manifold with corners, with $C_2(\Sigma)$ defined as above.

(2) Each projection $C_n^0(\Sigma) \to C_k^0(\Sigma)$, $k < n$, corresponding to a choice of $k$ points, extends to a smooth map $C_n(\Sigma) \to C_k(\Sigma)$ of manifolds with corners.

As the boundary of $C_n(\Sigma)$ has measure zero, we can replace the integration on $C_n^0(\Sigma)$ with an integration on $C_n(\Sigma)$. The integrand is now a smooth form obtained as a wedge product of pullbacks $\pi_{ij}^*\eta$ and $(\pi_i^*\underline{a}, \pi_i^*\underline{b}, \pi_i^*\mathcal{A}, \pi_i^*\mathcal{B})$. Here $\pi_{ij}\colon C_n(\Sigma) \to C_2(\Sigma)$ is the projection corresponding to a pair of vertices $i$ and $j$ with an $\widetilde{a}$ on the first and a $\widetilde{b}$ on the second joined to produce a propagator; $\pi_i\colon C_n(\Sigma) \to C_1(\Sigma) = \Sigma$ is the projection corresponding to a vertex $i$ where one of the forms $(\underline{a}, \underline{b}, \mathcal{A}, \mathcal{B})$ is placed.

As we are now left with the integral of a smooth form on a compact manifold with boundary, convergence of the integral is ensured.

Moreover, there is a standard technique to obtain the properties of the Feynman expansion, showing that we really get an observable for the ambient $BF$ theory coupled to the residual fields $(\underline{a}, \underline{b})$, which uses Stokes theorem.

## 6. 2-KNOT INVARIANTS

In this section, we describe how to use the surface observables in $BF$ theory to obtain invariants of 2-knots.

Let $M$ be a closed four-manifold. A 2-knot in $M$ is an embedding $\sigma\colon \Sigma \to M$, with $\Sigma$ a closed two-manifold. The expectation value

$$I_\sigma \coloneqq \langle \mathcal{O}_\sigma \rangle$$

is then well-defined and invariant under isotopies (see Section 4.1).

To be precise, we have to remember that the perturbative construction of $\mathcal{O}_\sigma$ yields a half-density on the space of the residual fields for the 2D-$BF$-theory on $\Sigma$ parametrized by the ambient 4D-$BF$-theory fields and the 2D-$BF$-theory background fields (see Remark 5.1).

The perturbative evaluation of $I_\sigma$ is obtained by integrating out the fluctuations of the ambient-$BF$-theory fields around their own background and residual fields. At the end, we do not get a numerical invariant: for a fixed choice of background fields,[16] $I$ is a map from the space $K(\Sigma, M)$ of embeddings $\Sigma \to M$ to the space of solutions of the quantum master equations on the BV space $\mathcal{F}_{\Sigma,\mathrm{res}} \times \mathcal{F}_{M,\mathrm{res}}$:

$$I\colon K(\Sigma, M) \to \mathcal{D}^{\frac{1}{2}}(\mathcal{F}_{\Sigma,\mathrm{res}} \times \mathcal{F}_{M,\mathrm{res}}), \qquad \triangle K_\sigma = 0 \ \forall \sigma \in K(\Sigma, M),$$

---

[16]One could further work in the direction of globalization relating $I$ to a globally well-defined density on the space of background fields, along the lines of [24, 14, 30]. We will not pursue this here.

where $\mathcal{D}^{\frac{1}{2}}$ denotes the space of half-densitites and $\triangle$ is the canonical BV Laplacian.[17]

*Remark* 6.1 (The invariant). Under an isotopy of $\sigma$, $I_\sigma$ changes by a $\triangle$-exact term. At the end of the day, the 2-knot invariant takes values in the quantum BV cohomology of $\mathcal{F}_{\Sigma,\text{res}} \times \mathcal{F}_{M,\text{res}}$.

*Remark* 6.2 (The trivial case). In the trivial case studied in the previous section, we have the identification

$$\mathcal{F}_{\Sigma,\text{res}} \times \mathcal{F}_{M,\text{res}} \simeq (H^\bullet(\Sigma)[1] \oplus H^\bullet(\Sigma))) \otimes \mathfrak{g} \times (H^\bullet(M)[1] \oplus H^\bullet(M)[2])) \otimes \mathfrak{g}.$$

6.1. **Long 2-knots.** A simpler situation where we can give more explicit formulae and get a numerical invariant is the one considered in [31] (only for the case $\lambda = 0$). Namely, we take $M = \mathbb{R}^4$, $\Sigma = \mathbb{R}^2$, and we fix a reference embedding $\sigma_0 \colon \mathbb{R}^2 \to \mathbb{R}^4$, e.g.,

$$\sigma_0(x, y) = (x, y, 0, 0).$$

A long 2-knot is an embedding $\sigma \colon \mathbb{R}^2 \to \mathbb{R}^4$ that outside a compact subset is equal to $\sigma_0$.

To compute invariants of long 2-knots, we resort to the expectation value of $\mathcal{O}_\sigma$ in $BF$ theory. Since now $M$ and $\Sigma$ are not closed manifolds, we have to impose appropriate conditions at infinity for the fields.

6.1.1. *The construction of $\mathcal{O}_\sigma$.* We start with a discussion of the 2D $BF$ theory. With the exception of $\beta$, we require all the BV fields to vanish at infinity. In order to get an interesting result, we instead require $\beta$ to converge at infinity to $\hbar \xi$, where $\xi$ is an element of the Lie algebra $\mathfrak{g}$ chosen once and for all (it becomes part of the definition of the observable). Accordingly, we choose the background fields as[18]

$$\alpha_0 = 0, \qquad \beta_0 = \hbar \xi, \qquad A_0 = 0, \qquad B_0 = 0.$$

---

[17]Recall (Remark 5.1) that $\mathcal{O}_\sigma$ is already a half-density on $\mathcal{F}_{\Sigma,\text{res}}$. The perturbative computation of the gauge-fixed BV integral $\langle \mathcal{O}_\sigma \rangle$ over the fluctuations of the ambient $BF$ theory yields a half-density on $\mathcal{F}_{M,\text{res}}$. This is the product of an analytic torsion, indeed a half-density, coming from the Berezinian of the quadratic part times the perturbative expansion in Feynman diagrams.

[18]We can make these choices up to a gauge transformation. For $\beta_0$, which is a constant map, there is no choice. The reason for the $\hbar$ factor, also present in [31], is that the term $\beta B$ in the $J$ action will contain a leading term $\hbar \xi \int_\Sigma B$. In the resulting observable, we will then have a leading factor $\exp i\xi \int_\Sigma B$. This is consistent with the usual definition of a Wilson loop as $\text{tr}_\rho P \exp iA$ with no $\hbar$ factor.

The decomposition (5.1) then becomes

$$a = \widehat{a},$$
$$b = \hbar\xi + \widehat{b},$$
$$\mathcal{A} = \widehat{\mathcal{A}},$$
$$\mathcal{B} = \widehat{\mathcal{B}},$$

with all the hatted fields vanishing at infinity. In the topology under consideration, there are no residual fields, so, actually,

$$\widehat{a} = \widetilde{a}, \qquad \widehat{b} = \widetilde{b}, \qquad \widehat{\mathcal{A}} = \widetilde{\mathcal{A}}, \qquad \widehat{\mathcal{B}} = \widetilde{\mathcal{B}}.$$

In summary, we can rewrite the BV surface action as

$$\mathcal{J}_\xi[\widetilde{a}, \widetilde{b}; \widetilde{\mathcal{A}}, \widetilde{\mathcal{B}}]$$
$$= \int_{\mathbb{R}^2} \left( \widetilde{a}\mathrm{d}\widetilde{b} + \widetilde{a}[\widetilde{\mathcal{A}}, \widetilde{a}] + \frac{\lambda}{2}\widetilde{b}[\widetilde{a}, \widetilde{a}] + \widetilde{b}\widetilde{\mathcal{B}} + \hbar\xi[\widetilde{\mathcal{A}}, \widetilde{a}] + \frac{\lambda}{2}\hbar\xi[\widetilde{a}, \widetilde{a}] + \hbar\xi\widetilde{\mathcal{B}} \right).$$

From this we can read propagator and vertices for the surface theory.

We start with the propagator. The discussion is as in Section 5.1 with some simplifications. The first is that there are no residual fields. The second is that we can give a very explicit formula for the propagator kernel $\eta$ of Section 5.1.2, analogously to [42, 9, 18]. Namely, we consider the map

$$\begin{array}{rcl} \phi\colon C_2^0(\mathbb{R}^2) & \to & S^1 \\ (x, y) & \mapsto & \frac{x-y}{\|x-y\|} \end{array}$$

and get $\eta = \frac{1}{2\pi}\phi^*\mathrm{d}\vartheta$, with $\vartheta \in [0, 2\pi]$. In Feynman diagrams, this propagator, viewed as the expectation value $\langle \widetilde{a}\widetilde{b} \rangle$, will be denoted by ---▶---.

As for the vertices, first we have $\hbar\xi B$ with no propagators attached. Then we have the vertices independent of $\lambda$:

-▶- $\widetilde{\mathcal{A}}$ -▶- ,    $[\xi, \widetilde{\mathcal{A}}]$ ---▶-- ,    -▶- $\widetilde{\mathcal{B}}$ .

Finally, the vertices with $\lambda$ are

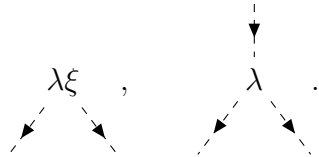

We may now present some Feynman diagrams contributing to $\mathcal{O}_\sigma$.[19] At $\lambda = 0$, we only have the diagrams studied in [31, 62]: the "snakes" of various length

$$\xi\widetilde{\mathcal{B}}\,, \qquad [\xi,\widetilde{\mathcal{A}}]\dashrightarrow\widetilde{\mathcal{B}}\,, \qquad [\xi,\widetilde{\mathcal{A}}]\dashrightarrow\widetilde{\mathcal{A}}\dashrightarrow\widetilde{\mathcal{B}}\,, \qquad \ldots,$$

and the "wheels," also of various length,

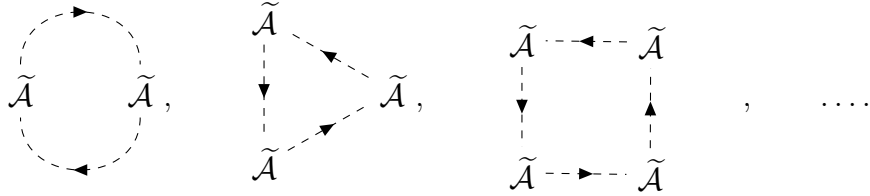

If $\lambda \neq 0$, we also have other diagrams. For esample, a snake might start bifurcating

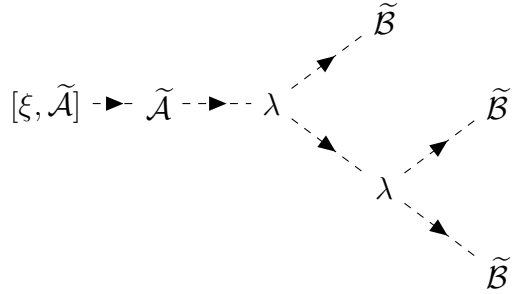

or a wheel might start growing branches

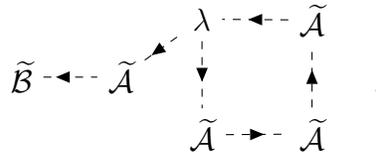

.

---

[19]The integrals are over $C_k(\mathbb{R}^2)$, the compactified configuration spaces of $\mathbb{R}^2$, where $k$ is the number of vertices. In this case, also compactification at infinity is involved. Concretely, $C_k(\mathbb{R}^2)$ may be defined as follows [18]. One first compactifies $\mathbb{R}^2$ to the sphere $S^2$, with $\mathbb{R}^2 = S^2 \setminus \{N\}$. Then one defines $C_k(\mathbb{R}^2)$ as $\pi_{k+1}^{-1}(N) \subset C_{k+1}(S^2)$.

Moreover, we will have diagrams starting at a $\lambda\xi$ vertex, e.g.,

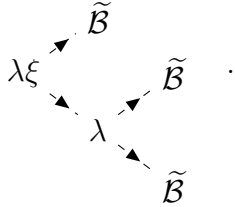

*Remark* 6.3. It is possible to show that the degree in $\hbar$ of the Feynman diagrams is equal to the sum of the degree in $\xi$ and the degree in $\lambda$:

$$\deg_\hbar = \deg_\xi + \deg_\lambda.$$

The only diagram of degre 1 in $\hbar$ is the vertex $\xi\widetilde{\mathcal{B}}$. Therefore,

$$\mathcal{O}_\sigma = e^{i\xi \int_{\mathbb{R}^2} \mathcal{B} + \cdots}. \tag{6.1}$$

Note that, thanks to the invariant pairing on $\mathfrak{g}$, we may also think of $\xi$ as an element of $\mathfrak{g}^*$ (this is actually the canonical choice in the case of pure $BF$ theory, i.e., when $\lambda = 0$). We may therefore think of $\xi$ as the analogue of choosing a representation of $\mathfrak{g}$ in the definition of a Wilson loop.

6.1.2. *The expectation value of $\mathcal{O}_\sigma$.* The BV action for the 4D $BF$ theory may now be written as

$$\mathcal{S} = \int_{\mathbb{R}^4} \left( \widetilde{B}\,d\widetilde{A} + \frac{1}{2}\widetilde{B}[\widetilde{A},\widetilde{A}] - \frac{\lambda}{2}\widetilde{B}\widetilde{B} \right).$$

The propagator can also be constructed analogously to [42, 9, 18]. Namely, we consider the map

$$\begin{array}{rcl} \Phi\colon C_2^0(\mathbb{R}^4) & \to & S^3 \\ (x,y) & \mapsto & \frac{x-y}{\|x-y\|} \end{array}$$

and get the propagator $\theta = \Phi^*\omega$, where $\omega$ is the normalized, uniform volume form on $S^3$. In Feynman diagrams, this propagator, viewed as the expectation value $\langle\widetilde{\mathcal{B}}\widetilde{\mathcal{A}}\rangle$, will be denoted by ⟶. We have two vertices:

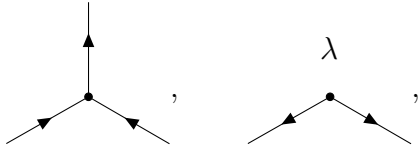

where the bullet are to remember that the vertices are in $\mathbb{R}^4$ now.

In the expectation value of $\mathcal{O}_\sigma$, we will then have two type of vertices, corresponding to integration on $\mathbb{R}^4$ or on $\mathbb{R}^2$, and two type of propagators.[20]

The lowest order in $\xi$ (second order) and no $\lambda$ is as in [31]. It is the sum of the the three following diagrams.

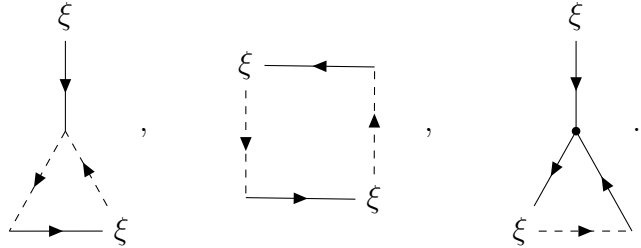

This is also the same as the 2-knot invariant proposed by Bott [16]. It is been proved that the $\lambda$-independent diagrams produce invariants related to the coefficients of the Alexander invariant [68, 49, 50]. These invariants should also be related to those constructed in [10].

Switching on $\lambda$ might produce new independent invariants. We now have several new possibilities. For example, here is a diagram at order $\lambda^2$ independent of $\xi$:

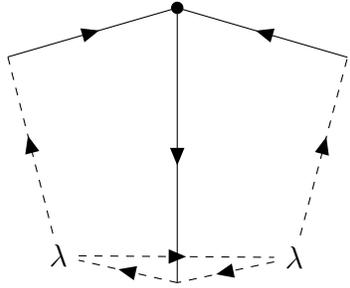

## 7. BV surface observables for Yang–Mills theory

As discovered in [13], 4D (euclidean) Yang–Mills theory can be recovered from $BF$ theory with cosmological term via a BV pushforward. This will allow us turning our observable $\mathcal{O}_\sigma$ into a surface observable for Yang–Mills theory.

---

[20]The compactified configuration spaces $C_k(\mathbb{R}^4)$ are defined along the lines of footnote 19. The embedding $\sigma$ then extends to a smooth map $C_k(\mathbb{R}^2) \to C_k(\mathbb{R}^4)$. For every $n$, we also have the map $C_{k+n}(\mathbb{R}^4) \to C_k(\mathbb{R}^4)$ that forgets the last $n$ points. The compactified configuration space $C_{k,n}$ of $n+k$ points in $\mathbb{R}^4$, $k$ of which lie in the image of $\sigma$, is defined [18] as the fibered product of $C_k(\mathbb{R}^2)$ and $C_{k+n}(\mathbb{R}^4)$ over $C_k(\mathbb{R}^4)$.

7.1. **Yang–Mills theory from $BF$ theory.** Let us recall the construction of [13]. We start with the BV action (4.2) for $BF$ theory. Next we pick the riemannian metric we want to use to define euclidean Yang–Mills theory and split the field $B$ and its antifield $B^+$ into their self-dual and anti-self-dual components:[21]

$$B = B_+ + B_-, \qquad B^+ = B^+_+ + B^+_-.$$

Then we split the space $\mathcal{F}$ of BV fields for $BF$ theory into the product $\mathcal{F}_1 \times \mathcal{F}_2$, where, in the notation of (4.1), we use $(c, A, B^+_+, B_+, A^+, c^+)$ as coordinates for $\mathcal{F}_1$ and $(B^+_-, \tau^+, \phi^+, \phi, \tau, B_-)$ as coordinates for $\mathcal{F}_2$. If we split the original BV action (4.2),

$$\mathcal{S} = \int_M \left( BF_A - \frac{\lambda}{2} BB + A^+(\mathrm{d}_A c - \lambda \tau) + B^+([c, B] + \mathrm{d}_A \tau) \right.$$
$$\left. + c^+ \left( \frac{1}{2}[c,c] - \lambda \phi \right) + \tau^+([c, \tau] + \mathrm{d}_A \phi) + \phi^+[c, \phi] + \frac{1}{2}[B^+, B^+] \right),$$

as $\mathcal{S}^{1\text{st}}_{\text{YM}} + \mathcal{S}_2$ with

$$\mathcal{S}^{1\text{st}}_{\text{YM}} = \int_M \left( B_+ F_A - \frac{\lambda}{2} B_+ B_+ + A^+ \mathrm{d}_A c + B^+_+[c, B_+] + c^+ \frac{1}{2}[c,c] \right),$$

we recognize $(\mathcal{F}_1, \omega_1, \mathcal{S}^{1\text{st}}_{\text{YM}})$ as the BV triple for Yang–Mills theory in the first-order formalism [23, 22, 34].

**Proposition 7.1.** *Ignoring residual fields, there is a choice of gauge-fixing Lagrangian for which the BV pushforward of $\mathcal{S}$ to $\mathcal{F}_1$ yields the effective action $\mathcal{S}_1 = \mathcal{S}^{1st}_{YM} + \hbar \mathcal{S}_{corr}$.*

*Remark* 7.2. For this computation we work around a flat reference connection $A_0$. The residual fields then belong to the cohomology of the covariant exterior derivative $\mathrm{d}_{A_0}$. A case of particular interest where there are no residual fields is that of $M = \mathbb{R}^4$ and $A_0 = 0$—the trivial connection. If residual fields are present, they will give contribution of order 1 to the effective action. One can set them to zero or force them to zero by inserting appropriate observables. We will not discuss this in this paper.

*Remark* 7.3. The quantum corrections in $\mathcal{S}_{\text{corr}}$ are nonlocal and depend on the chosen riemannian metric. We may view $\mathcal{S}_{\text{corr}}$ as a cloud correcting Yang–Mills theory at the quantum level. In this version, the surface observable will be obtained directly from $BF$ theory via BV pushforward. Alternatively, we may prefer to work with standard Yang–Mills theory and use the cloud $\mathcal{S}_{\text{corr}}$ to correct the surface observable. See

---

[21]To work in lorentzian signature, we should use complex-valued fields.

Section 7.2 for the full discussion. Clouds appearing in field theory and affecting observables have appeared in a different context, also related to BV pushforward, in [4].

*Proof of Proposition 7.1.* What we have to show is that the Feynman diagrams in the BV pushforward contain no trees and only one-loop diagrams if the residual fields are set to zero.

For the BV pushforward from $\mathcal{F}$ to $\mathcal{F}_1$, we first need the explicit form of $\mathcal{S}_2$:

$$\mathcal{S}_2 = \int_M \left( B_- F_A - \frac{\lambda}{2} B_- B_- + \lambda A^+ \tau + B_+^* d_A \tau + B_-^+([c, B_-] + d_A \tau) \right.$$
$$\left. + \lambda c^+ \phi + \tau^+([c, \tau] + d_A \phi) + \phi^+[c, \phi] + \frac{1}{2}\phi[B_-^+, B_-^+] + \frac{1}{2}\phi[B_+^+, B_+^+] \right).$$

To perform the BV pushforward over the variables $(B_-^+, \tau^+, \phi^+, \phi, \tau, B_-)$,[22] we take the gauge fixing

$$\phi^+ = 0, \qquad B_- = 0, \qquad \tau^+ = \tau_{\mathrm{gf}}^+ \in \mathrm{im}\, d_{A_0}^*, \qquad \tau = \tau_{\mathrm{gf}} \in \mathrm{im}\, d_{A_0}^*,$$

where $d_{A_0}^*$ is the formal adjoint of the covariant exterior derivative $d_{A_0}$ relative to the background connection $A_0$ we are working around. The gauge-fixed action becomes

$$(7.1) \quad \mathcal{S}_2^{\mathrm{gf}} = \int_M \left( \lambda A^+ \tau_{\mathrm{gf}} + B_+^* d_A \tau_{\mathrm{gf}} + B_-^+ d_A \tau_{\mathrm{gf}} + \lambda c^+ \phi \right.$$
$$\left. + \tau_{\mathrm{gf}}^+([c, \tau_{\mathrm{gf}}] + d_A \phi) + \frac{1}{2}\phi[B_-^+, B_-^+] + \frac{1}{2}\phi[B_+^+, B_+^+] \right).$$

A very simple observation now is that rescaling $\phi \mapsto \hbar\phi$ and $\tau \mapsto \hbar\tau$ rescales $\mathcal{S}_2^{\mathrm{gf}}$ by $\hbar$. As a result, $e^{\frac{i}{\hbar}\mathcal{S}_2^{\mathrm{gf}}}$ becomes $e^{i\mathcal{S}_2^{\mathrm{gf}}}$, which is completely independent of $\hbar$. It follows that its BV pushforward, which we denote by $e^{i\mathcal{S}_{\mathrm{corr}}}$ is also independent of $\hbar$. □

*Remark* 7.4. We can interpret the correction term $e^{i\mathcal{S}_{\mathrm{corr}}}$ as a modification to the naive functional measure of Yang–Mills theory.

*Remark* 7.5 (Gauge transformations). Since $\mathcal{S}_2^{\mathrm{gf}}$ depends on $A^+$, the gauge transformations for $A$ in $\mathcal{S}_1$ are deformed by a term of order $\hbar$.

---

[22]To be more precise, we should first select background and residual fields. For simplicity, also of notation, we ignore this. Note that, working on $M = \mathbb{R}^4$, we may assume background and residual fields to vanish.

7.1.1. *More details on the BV pushforward.* It may also be useful to see more in details what kind of Feynman diagrams are generated by the BV pushforward of $\mathcal{S}_2^{\text{gf}}$. We write $\mathcal{S}_2^{\text{gf}} = \mathcal{S}_{\text{kin}} + \mathcal{S}_{\text{int}} + \mathcal{S}_{\text{sou}}$ with

$$\mathcal{S}_{\text{kin}} = \int_M \left( B_-^+ \mathrm{d}_{A_0} \tau_{\text{gf}} + \tau_{\text{gf}}^+ \mathrm{d}_{A_0} \phi \right),$$

$$\mathcal{S}_{\text{int}} = \frac{1}{2} \int_M \phi[B_-^+, B_-^+],$$

$$\mathcal{S}_{\text{sou}} = \int_M \left( \lambda A^+ \tau_{\text{gf}} + B_+^+ \mathrm{d}_{A_0} \tau_{\text{gf}} + B_+^+ [\widehat{A}, \tau_{\text{gf}}] + B_-^+ [\widehat{A}, \tau_{\text{gf}}] + \lambda c^+ \phi \right.$$
$$\left. + \tau_{\text{gf}}^+ [c, \tau_{\text{gf}}] + \tau_{\text{gf}}^+ [\widehat{A}, \phi] + \frac{1}{2} \phi[B_+^+, B_+^+] \right)$$

with $\widehat{A} = A - A_0$.

The terms in $\mathcal{S}_{\text{kin}}$ produce a determinant (of order 1 and hence a correction of oder $\hbar$ to the effective action) and the propagators $\langle B_-^+ \, \tau_{\text{gf}} \rangle$ and $\langle \tau_{\text{gf}}^+ \, \phi \rangle$, both of order $\hbar$. The vertices $B_-^+[\widehat{A}, \tau_{\text{gf}}]$ and $\tau_{\text{gf}}^+[\widehat{A}, \phi]$ may be used to produce long strings, which we may call fat propagators $\ll B_-^+ \, \tau_{\text{gf}} \gg$ and $\ll \tau_{\text{gf}}^+ \, \phi \gg$, both again of order $\hbar$.

Next we look at the resulting diagrams, which will produce corrections to $\mathcal{S}_{\text{YM}}^{\text{1st}}$ in the effective action. One class of diagrams consists in closing a fat propagator, joining the initial with the final vertex. This produces a wheel. As each wheel has order 1, the correction to the effective action will be of order $\hbar$.

The remarkable point is that there are no diagrams at order $1/\hbar$ (so no correction to the effective action at order 1). In fact, we are left with terms linear in $\tau_{\text{gf}}$, terms linear in $\phi$, and the term $\tau_{\text{gf}}^+[c, \tau_{\text{gf}}]$ which is bilinear in $\tau_{\text{gf}}$ and $\tau_{\text{gf}}^+$. Moreover, the only interaction term is in $\mathcal{S}_{\text{int}}$. There is no way to construct a tree connecting these vertices by fat propagators.

Thus, all the possible diagrams have at least one loop, which means they are at least at order 1, providing corrections to the effective action of order at least $\hbar$. However, from the combinatorics of the vertices, it is clear we cannot produce more than one loop.

7.1.2. *From first to second order.* The second-order formulation of Yang–Mills theory is easily recovered from $\mathcal{S}_{\text{YM}}^{\text{1st}}$ by a further BV pushforward over $(B_+, B_+^+)$. As a gauge-fixing we choose $B_+^+ = 0$. The gaussian integration over $B_+$ then yields

$$\mathcal{S}_{\text{YM}} = -\int_M \left( \frac{1}{2\lambda} (F_A)_+ (F_A)_+ - A^+ \mathrm{d}_A c - c^+ \frac{1}{2}[c, c] \right)$$

This is the BV action for Yang–Mills theory with a specific choice of topological term.[23] Starting from $\mathcal{S}^\kappa$ as defined in Section 4.2.1 yields instead
$$\int_M \left(\frac{1}{4\lambda}F_A * F_A + \left(\frac{1}{4\lambda} + \frac{\kappa}{2}\right) F_A F_A - A^+ \mathrm{d}_A c - c^+ \frac{1}{2}[c,c]\right),$$
and we can get whatever theta angle we prefer.

Since we start from the modified action $\mathcal{S}_1 = \mathcal{S}_{\mathrm{YM}}^{\mathrm{1st}} + \hbar \mathcal{S}_{\mathrm{corr}}$, the BV pushforward considered here will produce corrections to $\mathcal{S}_{\mathrm{YM}}$ at the first order in $\hbar$.

7.1.3. *Wilson loops.* Wilson loops are observables for Yang–Mills theory, both in the first- and in the second-order formalism. They are not observables for the modified theory because the gauge transformations for $A$ are changed, see Remark 7.5.

We can remedy for this by dressing the Wilson loop $W_\gamma$ associated to a loop $\gamma$ (and some representation) to get the BV observable
$$\mathcal{W}_\gamma := \mathrm{e}^{-\mathrm{i}\mathcal{S}_{\mathrm{corr}}} W_\gamma$$
for the BV theory defined by $\mathcal{S}_1$.

Note that while $W_\gamma$ only depends on $A$, the BV observable $\mathcal{W}_\gamma$ depends on all the BV fields $(c, A, B_+^+, B_+, A^+, c^+)$. Moreover, $\mathcal{W}_\gamma$ is no longer localized on $\gamma$ because the correction term $\mathcal{S}_{\mathrm{corr}}$ is highly nonlocal.

7.2. **Surface observables.** To get surface observables for Yang–Mills theory, we start from the $BF$ theory action $\mathcal{S}$ and from the surface action $\mathcal{J}$. As we know from Section 4, we can consider the sum $\widehat{\mathcal{S}} := \mathcal{S} + \mathcal{J}$ as a BV action. If we then perform the BV pushforward jointly over $\mathcal{F}_2$, as in the last subsection, and over the surface fields $a$ and $b$, we get an observable $\mathcal{O}_\sigma^{\mathrm{YM},\mathrm{1st}}$ for Yang–Mills theory plus the $\hbar$-correction. We might also decide to integrate only over $\mathcal{F}_2$. In this case, the resulting Yang–Mills observable will still be expressed in terms of a functional integral over surface fields.

To obtain an observable for the standard first-order Yang–Mills theory $\mathcal{S}_{\mathrm{YM}}^{\mathrm{1st}}$, we may dress $\mathcal{O}_\sigma^{\mathrm{YM},\mathrm{1st}}$ with the correction term. Namely, we define
$$\widehat{\mathcal{O}}_\sigma^{\mathrm{YM},\mathrm{1st}} := \mathrm{e}^{\mathrm{i}\mathcal{S}_{\mathrm{corr}}} \mathcal{O}_\sigma^{\mathrm{YM},\mathrm{1st}},$$
which is now a quantum BV observable for $\mathcal{S}_{\mathrm{YM}}^{\mathrm{1st}}$. Note that the price we pay is that $\widehat{\mathcal{O}}_\sigma^{\mathrm{YM},\mathrm{1st}}$ is no longer localized on the image of $\sigma$.

---

[23]The weight in the resulting functional integral is $\exp\left(\frac{\mathrm{i}}{\hbar}\mathcal{S}_{\mathrm{YM}}\right)$ even though we are in euclidean signature. We should see this as an analytic continuation of the euclidean functional integral for purely imaginary $\hbar\lambda$.

The expectation value of $\mathcal{O}_\sigma^{\text{YM,1st}}$ in Yang–Mills theory plus the $\hbar$-correction (or of $\widehat{\mathcal{O}}_\sigma^{\text{YM,1st}}$ in the standard first-order Yang–Mills theory) is, by construction, the same as the expectation value of $\mathcal{O}_\sigma$ in $BF$ theory, so an invariant of the embedding $\sigma$.

*Remark* 7.6 (Expectation values). To get nontopological results, we should compute the expectation value of $\mathcal{O}_\sigma^{\text{YM,1st}}$ (possibly for several disjoint embeddings) multiplied by other Yang–Mills observables. Recall however that quantum BV observables do not form an algebra, since the BV Laplacian is a second-order operator. As a consequence the product of the surface observable and, e.g., a Wilson loop is not an observable.[24] Therefore, electric and magnetic fluxes cannot be measured simultaneously (although by a different mechanism than in the abelian case considered in [38, 63]).

Our last remark is that, under the assumptions of Section 6.1, see equation (6.1), the Yang–Mills observable will have the form

$$\mathcal{O}_\sigma^{\text{YM,1st}} = e^{i \int_\Sigma \xi B_+ + \cdots}.$$

If we then go to the second-order formulation, as in Section 7.1.2, we get

(7.2) $$\mathcal{O}_\sigma^{\text{YM,2nd}} = e^{\frac{i}{\lambda} \int_\Sigma \xi (F_A)_+ + \cdots} = e^{\frac{i}{2\lambda} \int_\Sigma \xi (*F_A + F_A) + \cdots}.$$

This is a mixture of electric and magnetic observables. We propose two ways to recover a purely electric observable.

7.2.1. *Electric observable I: Change the initial observable.* Instead of $\mathcal{J}$, we may consider the observable $\mathcal{J}_\gamma$ defined in (4.4). In this case, the computations above yield

$$\mathcal{O}_\sigma^{\text{YM,2nd}} = e^{i(1-\lambda\gamma) \int_\Sigma \xi \left( \frac{(F_A)_+}{\lambda} - \gamma F_A \right) + \cdots} = e^{i(1-\lambda\gamma) \int_\Sigma \xi \left( \frac{*F_A}{2\lambda} + (\frac{1}{2\lambda} - \gamma) F_A \right) + \cdots},$$

so it is enough to choose $\gamma = \frac{1}{2\lambda}$ to get the purely electric observable

$$\mathcal{O}_\sigma^{\text{YM,2nd,electric}} = e^{\frac{i}{4\lambda} \int_\Sigma \xi * F_A + \cdots}.$$

Note that to achieve this we have to start with

$$\mathcal{J}_{\frac{1}{2\lambda}} = \frac{1}{2} \int_\Sigma \left( b\, d_\mathcal{A} a + \lambda b[a,a] + b\mathcal{B} - \frac{1}{2\lambda} b F_\mathcal{A} \right).$$

---

[24]A different strategy would consist in finding first an observable $\widehat{W}_\gamma$ for $BF$ theory whose BV pushforward is $W_\gamma$. The problem is that this observable is also highly nonlocal, so the product $\widehat{W}_\gamma \mathcal{O}_\sigma$ will likely also not be an observable.

By the rescaling $b \mapsto \lambda b$, $a \mapsto a/\lambda$, we can convert this to
$$\mathcal{J}_{\frac{1}{2\lambda}} = \frac{1}{2}\int_\Sigma \left( b\,\mathrm{d}_\mathcal{A} a + b[a,a] + \lambda b\mathcal{B} - \frac{1}{2}bF_\mathcal{A}\right).$$

7.2.2. *Electric observable II: Divide by a Wilson surface.* In [2] a functional-integral Wilson-surface observable for Yang–Mills theory was introduced (motivated by the path-integral presentation of Wilson loops of [3] via the nonabelian Stokes theorem of [36]). At the classical level, the authors also take an embedding $\sigma$ with image $\Sigma$ and consider the action

(7.3) $\quad Y_\sigma[\alpha,\beta;A] := \int_\Sigma \beta F_{A+\alpha} = \int_\Sigma \left(\beta\,\mathrm{d}_A\alpha + \frac{1}{2}\beta[\alpha,\alpha] + \beta F_A\right).$

We now proceed as in (2.2) and observe that the infinitesimal gauge transformation $\delta A = -\mathrm{d}_A\gamma$ can be compensated by the infinitesimal change of variables
$$\alpha \rightsquigarrow \alpha - [\gamma,\alpha], \quad \beta \rightsquigarrow \beta - [\gamma,\beta].$$
Extending this to the BV formalism and computing the BV pushforward over the surface fields yields an observable for Yang–Mills theory.

The problem now is that this is not an observable for the Yang–Mills theory with $\hbar$-correction $\mathcal{S}_1$ we obtained by BV pushforward, see Remark 7.5.

However, we may consider $Y_\sigma$ in (7.3) directly in $BF$ theory. It is easy to see that the infinitesimal symmetry transformations (2.3) are compensated by the infinitesimal change of variables
$$\alpha \rightsquigarrow \alpha + \lambda\theta - [\gamma,\alpha], \quad \beta \rightsquigarrow \beta - [\gamma,\beta].$$
Extending this to the BV formalism leads to the BV action
$$\mathcal{Y}_\sigma[a,b;\mathcal{A},\mathcal{B}] := \int_\Sigma bF_{\mathcal{A}+a} = \int_\Sigma \left(b\,\mathrm{d}_\mathcal{A} a + \frac{1}{2}b[a,a] + bF_\mathcal{A}\right).$$
Computing the BV pushforward over the surface fields yields an observable $\mathcal{U}_\sigma$ for $BF$ theory. Performing the BV pushforward as in Section 7.1 finally produces observables $\mathcal{U}_\sigma^{\mathrm{YM,1st}}$ and $\mathcal{U}_\sigma^{\mathrm{YM,2nd}}$ for Yang–Mills theory in the first- and second-order formalism plus the $\hbar$ corrections. Denoting by $\hbar\eta$ the background solution for $b$, and proceeding as in Section 7.2, we get
$$\mathcal{U}_\sigma^{\mathrm{YM,2nd}} = \mathrm{e}^{\mathrm{i}\int_\Sigma \eta F_A + \cdots}.$$

The rough idea is now to multiply the two $\mathcal{S}_1$-observables $\mathcal{O}_\sigma$ and $\mathcal{U}_\sigma$ to get rid of the magnetic part. Comparing with (7.2), we see that we should choose $\eta = -\frac{1}{2\lambda}\xi$ to achieve this. This idea does not work properly because the product of two quantum BV observables is not necessarily a quantum BV observable, as the BV Laplacian is a second-order differential operator.

This problem is however readily removed by a framing of $\Sigma$, namely, by displacing the surface $\Sigma$ in $\mathcal{Y}$ by some nowhere vanishing vector field $\epsilon Z$ transversal to $\Sigma$. The product of $e^{\frac{i}{\hbar}\mathcal{J}_\sigma}$ and $e^{\frac{i}{\hbar}\mathcal{Y}_{\sigma'}}$ is now a quantum BV observable, and its BV pushforward (with background fields $\xi$ and $\eta = -\frac{1}{2\lambda}\xi$) will produce, in the limit $\epsilon \to 0$ (to be taken after taking the expectation value), the desired purely electric observable for Yang–Mills theory.

## 8. Self-dual Yang–Mills theory

A simpler version of Yang–Mills theory, which fits very well in our description, is self-dual Yang–Mills theory [37, 61, 32, 52].

The BV space of fields $\mathcal{F}_1$ is the same as that for first-order Yang–Mills theory and has coordinates $(c, A, B_+^+, B_+, A^+, c^+)$. The BV action is instead

$$\mathcal{S}_{YM}^{sd} = \int_M \left( B_+ F_A + A^+ d_A c + B_+^+[c, B_+] + c^+ \frac{1}{2}[c, c] \right),$$

which is obtained from first-order Yang–Mills theory by setting $\lambda = 0$.

All the considerations of the previous sections now apply (with some simplifcation) to the present case. We just have to start from pure $BF$ theory and its corresponding surface observable, determined by $\mathcal{J}_\sigma$ with $\lambda = 0$. We have the analogue of Proposition 7.1.

**Proposition 8.1.** *Ignoring residual fields, there is a choice of gauge-fixing Lagrangian for which the BV pushforward of $\mathcal{S}$, with $\lambda = 0$, to $\mathcal{F}_1$ yields the effective action $\mathcal{S}_1 = \mathcal{S}_{YM}^{sd} + \hbar \mathcal{S}_{corr}$, and the surface observable $\mathcal{O}_\sigma$ for pure $BF$ theory determines an observable $\mathcal{S}_{YM}^{sd}$. In turn, the dressing*

$$\widehat{\mathcal{O}}_\sigma^{YM,sd} := e^{i\mathcal{S}_{corr}} \mathcal{O}_\sigma^{YM,sd}$$

*is an observable for self-dual Yang–Mills theory, with BV action $\mathcal{S}_{YM}^{sd}$.*

The difference is that now $A^+$ does not appear in $\mathcal{S}_2$, see (7.1), so neither in $\mathcal{S}_{\text{corr}}$, which means that the gauge transformations for $A$ in $\mathcal{S}_1$ are the usual ones, determined by $\mathcal{S}_{YM}^{sd}$. As a consequence, not only $\mathcal{O}_\sigma^{YM,sd}$ is an observable but also a Wilson loop $W_\gamma$ is so.

Since $W_\gamma$ only depends on $A$, which is a field in $\mathcal{F}_1$, and since it is an observable also for pure $BF$ theory, we can view it as a BV pushforward of itself.

If $\gamma$ and the image of $\sigma$ do not intersect, the product $\mathcal{O}_\sigma W_\gamma$ is a quantum BV observable for pure $BF$ theory, and its BV pushforward is the quantum BV obervable $\mathcal{O}_\sigma^{\text{YM,sd}} W_\gamma$ in the modified self-dual Yang–Mills theory determined by $\mathcal{S}_1$.

Finally, $\widehat{\mathcal{O}}_\sigma^{\text{YM,sd}} W_\gamma$ is a quantum BV observable for self-dual Yang–Mills theory, and its expectation value is a topological invariant.

Institut für Mathematik, Universität Zürich, Winterthurerstrasse 190, CH-8057 Zürich, Switzerland

*Email address*: cattaneo@math.uzh.ch